\documentclass[useAMS,usenatbib]{mn2e}
\usepackage{graphicx}
\usepackage{units}
\usepackage{amsmath}
\usepackage{pdfpages}
\usepackage[toc,page]{appendix}
\usepackage{amsfonts}
\usepackage{amssymb}
\usepackage{mathrsfs}
\usepackage{latexsym}
\usepackage{graphics}
\usepackage{cases}
\usepackage{mathtools}
\usepackage{enumitem}
\usepackage{myitemize}

\title[Extended Tully-Fisher Relations using {\HI} Stacking]{Extended Tully-Fisher Relations using {H{\,\Large I}} Stacking}
\author[Scott A. Meyer, Martin Meyer, Danail Obreschkow, Lister Staveley-Smith]{Scott A. Meyer$^{1,2}$\thanks{E-mail:
scott.meyer@icrar.org}, Martin Meyer$^{1,2}$, Danail Obreschkow$^{1,2}$ and Lister Staveley-Smith$^{1,2}$\\
$^{1}$International Centre for Radio Astronomy Research (ICRAR), University of Western Australia, 35 Stirling Hwy, Crawley, 6009, Australia\\
$^{2}$ARC Centre of Excellence for All-sky Astrophysics (CAASTRO)}

\DeclareRobustCommand{\ion}[2]{
\relax\ifmmode
\ifx\testbx\f@series
{\mathbf{#1\,\mathsc{#2}}}\else
{\mathrm{#1\,\mathsc{#2}}}\fi
\else\textup{#1\,{\mdseries\textsc{#2}}}%
\fi}
\makeatletter
\def\afterfi#1\fi{\fi#1}

\newcommand{\HI}{\textsc{H\,i}}
\newcommand{\vmax}{V_{\rm max}}

\newcommand{\be}{\begin{equation}}
\newcommand{\ee}{\end{equation}}
\newcommand{\bes}{\begin{equation}\begin{split}}
\newcommand{\ees}{{\end{split}\end{equation}}}

\renewcommand{\S}{Section~}

\begin{document}

\date{27 October 2015}

\pagerange{\pageref{firstpage}--\pageref{lastpage}} \pubyear{2013}

\maketitle

\label{firstpage}

\begin{abstract}
We present a new technique for the statistical evaluation of the Tully-Fisher relation (TFR) using spectral line stacking. This technique has the potential to extend TFR observations to lower masses and higher redshifts than possible through a galaxy-by-galaxy analysis. It further avoids the need for individual galaxy inclination measurements.

To quantify the properties of stacked {\HI} emission lines, we consider a simplistic model of galactic disks with analytically expressible line profiles. Using this model, we compare the widths of stacked profiles with those of individual galaxies. We then follow the same procedure using more realistic mock galaxies drawn from the S$^3$-SAX model (a derivative of the Millennium simulation). Remarkably, when stacking the {\it apparent} {\HI} lines of galaxies with similar absolute magnitude and random inclinations, the width of the stack is very similar to the width of the {\it deprojected} (=~corrected for inclination) and {\it dedispersed} (=~after removal of velocity dispersion) input lines. Therefore, the ratio between the widths of the stack and the deprojected/dedispersed input lines is approximately constant -- about 0.93 -- with very little dependence on the gas dispersion, galaxy mass, galaxy morphology, and shape of the rotation curve.

Finally, we apply our technique to construct a stacked TFR using HIPASS data which already has a well defined TFR based on individual detections. We obtain a B-band TFR with a slope of $-8.5 \pm 0.4$ and a K-band relation with a slope of $-11.7 \pm 0.6$ for the HIPASS data set which is consistent with the existing results.
\end{abstract}


\section{Introduction}

Galaxy scaling relations form an important part of extragalactic astrophysics as they encode the different evolutionary processes experienced by galaxies, and often serve as important observational tools. The Tully-Fisher relation (TFR), an empirical relation between the absolute magnitude and rotation velocity of spiral galaxies, is one of the most important of these as it links their dark and luminous matter components, as well as providing an important distance estimator in cosmology \citep{Tully1977,Springob2007,Masters2008}. Observations of the evolution in the TFR can potentially discriminate between different evolutionary models of spiral galaxies \citep{Obreschkow2009a}. Although TFR studies can be done in optical \citep{Miller2012,Puech2008}, the most accurate method for determining rotational velocities for TFRs is through the direct detection of atomic hydrogen ({\HI}, 21\,cm rest-frame), since this gas probes larger galactocentric radii than optical light, hence better tracing the asymptotic velocity (if it exists) of the rotation curve. However, given the technological difficulty of detecting distant \HI\ -- the current distance record being at redshift $z = 0.2454$ \citep{Catinella2008} -- TFR studies using this method are still limited to the local Universe ($z<0.1$). For gas evolution studies, {\HI} stacking has been successful in extending the redshift range accessible by measuring a statistical signal \citep{Lah2007,Delhaize2013,Rhee2013}, a method we now explore for studying the TFR.

In this paper, we describe the technique for recovering the TFR using stacked {\HI} profiles. \S\ref{HI stacking} provides background information on the idea and benefits of {\HI} stacking. \S\ref{analytical galaxies} uses analytical {\HI} profiles to assess how the widths of stacked profiles are related to the widths of the input profiles. All of the profiles used in \S\ref{analytical galaxies} were generated from identical galaxies with different inclination projections. This relationship is further investigated in \S\ref{Simulated galaxies} using a distribution of {\it non identical} and more realistic mock galaxies from the S$^3$-SAX simulation \citep{Obreschkow2009a,Obreschkow2009b}, which allows us to model volume and sensitivity-limited survey scenarios. Finally in \S\ref{HIPASS analysis} we use our method to create a stacked TFR from real HIPASS data and compare it to the relation derived from individual detections by \citet{Meyer2008}.

\section[]{{H{\,\small I}}  stacking}
\label{HI stacking}
\subsection[]{Introduction to {H{\,\small I}} stacking}
\label{General idea of HI stacking}

Stacking consists of adding or averaging the rest-frame spectra of multiple galaxies to produce one `stacked spectrum'. This approach is particularly interesting if the individual spectra are too noisy to reveal features such as emission lines. Assuming Gaussian noise, stacking increases the signal-to-noise as $\sqrt{N}$. Given a sufficiently large input sample, stacking uncovers the otherwise hidden 21 cm profile and enables a statistical flux or {\HI} mass measurement. In the context of the TFR, stacking should allow a measurement of the average rotation velocity of galaxies that are too distant or low in mass to be detected directly in {\HI} emission given the flux limit of the observation. While it is clear that the flux (or mass for mass-averaged spectra) of a stack equals the average flux (mass) of the galaxies used in the stack, it is not obvious that the width of galaxy {\HI} profiles is conserved in stacking. For this work, all of our spectra are mass-averaged spectra.

One of the challenges of the stacking method is that it requires the input spectra in their {\it rest-frame}, so that emission lines from the same transition add up constructively. In the case of the mock galaxies discussed in \S\ref{analytical galaxies} and \S\ref{Simulated galaxies}, we naturally have access to rest-frame {\HI} spectra, since the intrinsic properties of the mock galaxies are known by construction. In the case of observed {\HI} spectra, such as in the HIPASS data, we must first shift all galaxy spectra back to their rest-frame by using their redshifts. Since we are only using the direct detections from the HIPASS catalogue, we get this redshift information directly from the 21 cm profile. Stacking is done by taking the mean of the individual spectra in each frequency bin.

\subsection[]{Using stacked {H{\,\small I}} lines for TF science}
\label{Using stacked HI lines for TF science}

The fundamental axis of the TFR represents the rotation velocity of galaxies. The precise meaning of this rotation velocity can be ambiguous, especially in galaxies with rotation curves that do not converge toward a constant velocity with increasing radius. In this paper, we will not enter a discussion on different definitions of rotation velocities (e.g. $V_{\rm max}$, $V_{\rm flat}$, $W_{50}/2$, $W_{20}/2$) and on how they relate to the circular velocity and the mass of the halo. Instead, we explore how {\HI} stacking can be used to recover a TFR, within a fixed definition of the rotation velocity. We chose to define this velocity as half the width $W_{50}$ (measured at 50\% of the peak flux) of the {\HI} emission line of a galaxy,
if this galaxy were seen {\it edge-on} and had {\it no dispersion} in the {\HI} gas. To measure this velocity from an observed {\HI} emission line, the line profile has to be dedispersed (=~removal of dispersion) and deprojected (=~corrected for inclination). We label the dedispersed and deprojected line width $W_{50}^{\rm ind}$. The velocity to be plotted on the TFR is then $W_{50}^{\rm ind}/2$. In the case of a sample of galaxies with similar intrinsic rotation curves, we define the global average line width, $W_{50}^{\rm ref}$, as the {\HI}-mass weighted geometric average of the individual values (more on this in \S\ref{Simulated galaxies})

The leading question of this paper is, how can a stacked {\HI} line be used to measure $W_{50}^{\rm ref}$ of the stacked galaxies? More explicitly, how does the 50-percentile width $W_{50}^{\rm stack}$ of the stacked line compare to $W_{50}^{\rm ref}$? This question is non-trivial, because {\HI} lines entering the stack are {\it not} corrected for dispersion and inclination. In fact, dispersion correction is impossible in the case of non-detected lines -- the typical scenario of stacking. Inclination corrections (applied by stretching the observed {\HI} profiles in the frequency axis) could in principle be applied given optical inclinations, but these are one of the largest sources of error in TFR studies, as evidenced by the frequent use of an inclination selection criteria \citep{Meyer2008,Barton2001,Lagattuta2013}, and, as shown in this work, inclination corrections can be bypassed while still maintaining accurate results. For more work in deriving TFRs without using inclinations, see \citet{Obreschkow2013b}. In conclusion, a leading challenge of this work is to calculate the ratio
\begin{equation}\label{eq F}
	\mathcal{F}\equiv\frac{W_{50}^{\rm stack}}{W_{50}^{\rm ref}}.
\end{equation}
When constructing the TFR from stacked lines, $W_{50}^{\rm stack}/2$ needs to be multiplied by $\mathcal{F}^{-1}$ to obtain the correct velocity in the TFR. Sections \ref{analytical galaxies} and \ref{Simulated galaxies} therefore focus on determining the value of $\mathcal{F}$ in increasingly realistic models.

\section[]{Stacking identical mock galaxies}
\label{analytical galaxies}

In this section, we consider the line widths $W_{50}^{\rm stack}$ of stacked emission lines composed of {\it identical, but differently inclined} model galaxies. In other words, all the {\HI} lines in the stack would look the same if seen edge-on. In this case, the line width $W_{50}^{\rm ref}$, which we want to recover for the TFR, is simply the dedispersed edge-on line width of the input galaxies. In the following we study the ratio $\mathcal{F}$ (equation~\ref{eq F}) in the case of a simplistic disk with constant velocity (\S\ref{constant rotation velocity galaxies}) and varying velocity (\S\ref{galaxies with rotation curves}) rotation, and then investigate the bias and uncertainty of $\mathcal{F}$ if the stack only consists of a small number of individual lines (\S\ref{low-number effects}).

\subsection{Disk galaxy with constant circular rotation velocity}
\label{constant rotation velocity galaxies}

\begin{figure*}
	\begin{centering}
	\includegraphics[scale=0.43]{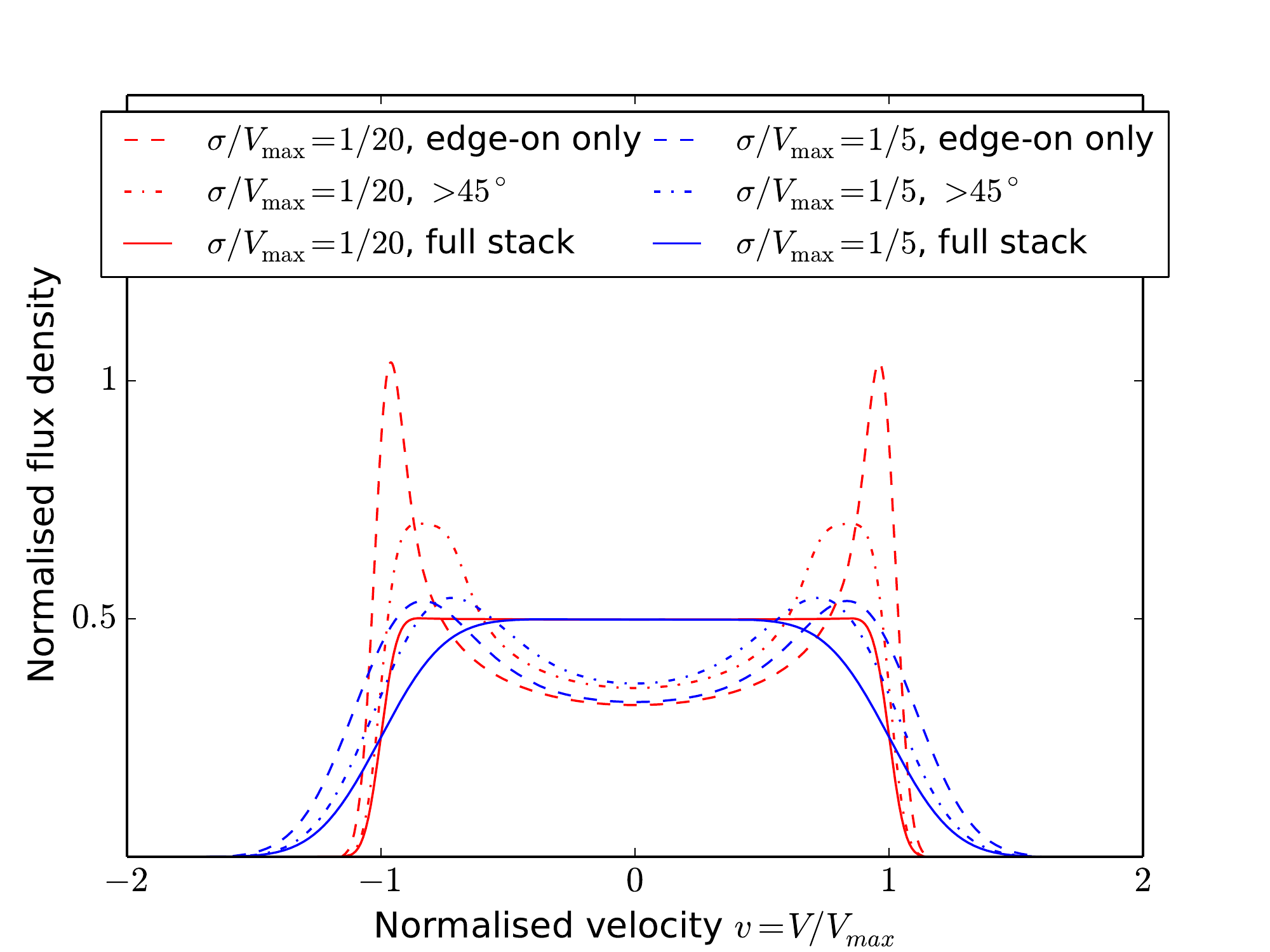}
	\caption{Example spectra generated using equations~(\ref{const rot}), (\ref{stackeqn}) and (\ref{add dispersion}) for galaxies with low dispersion (red) and high dispersion (blue). The dashed lines show spectra of edge-on galaxies. The dot-dashed lines show partial stacks containing galaxies between $45^{\circ}$ and $\mathrm{90^{\circ}}$. The solid lines show complete stacks containing galaxies of all inclinations.\label{Plot A1}}
	\par\end{centering}
\end{figure*}

\begin{figure*}
	\begin{centering}
	\includegraphics[scale=0.43]{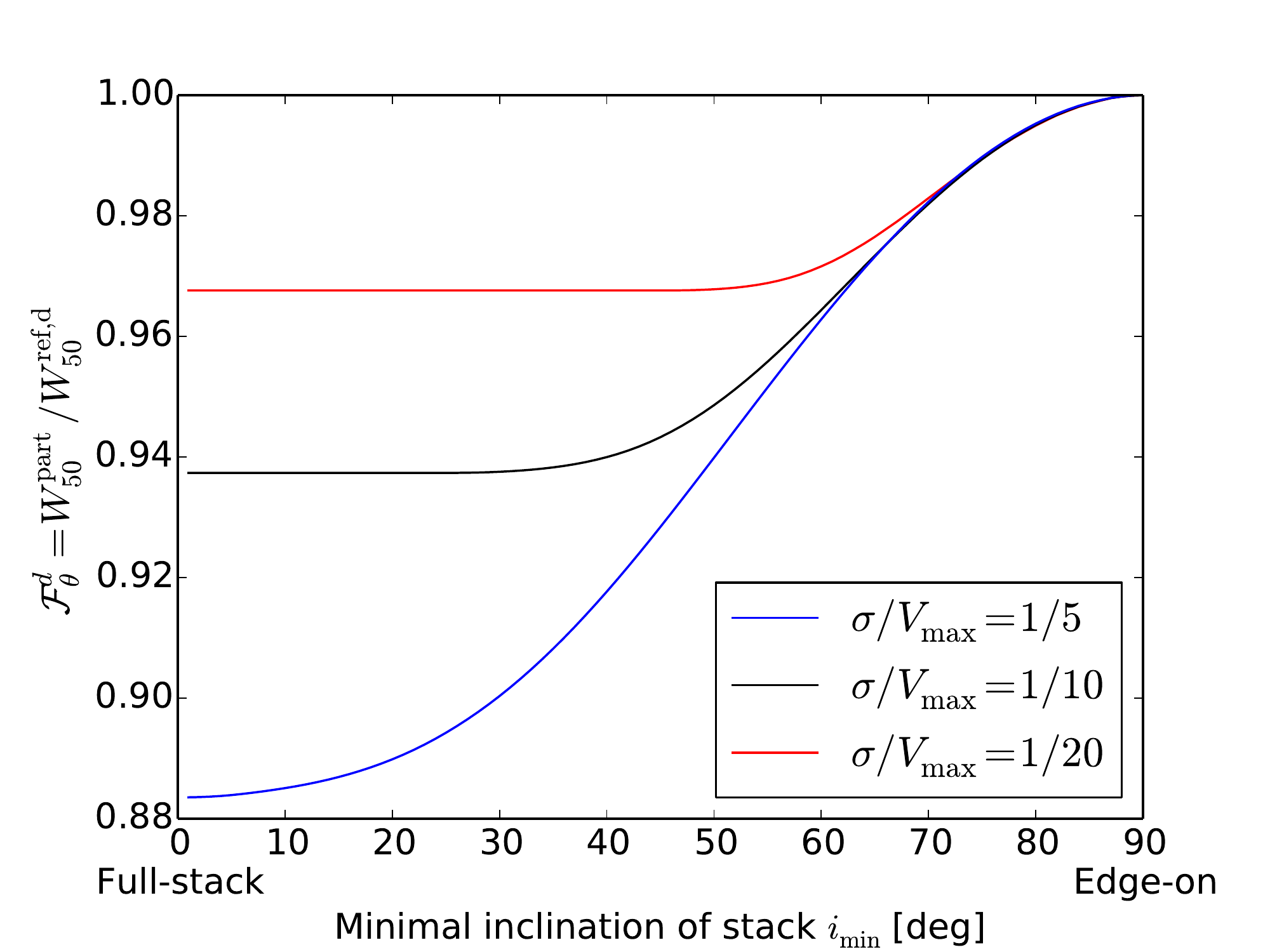}
	\includegraphics[scale=0.43]{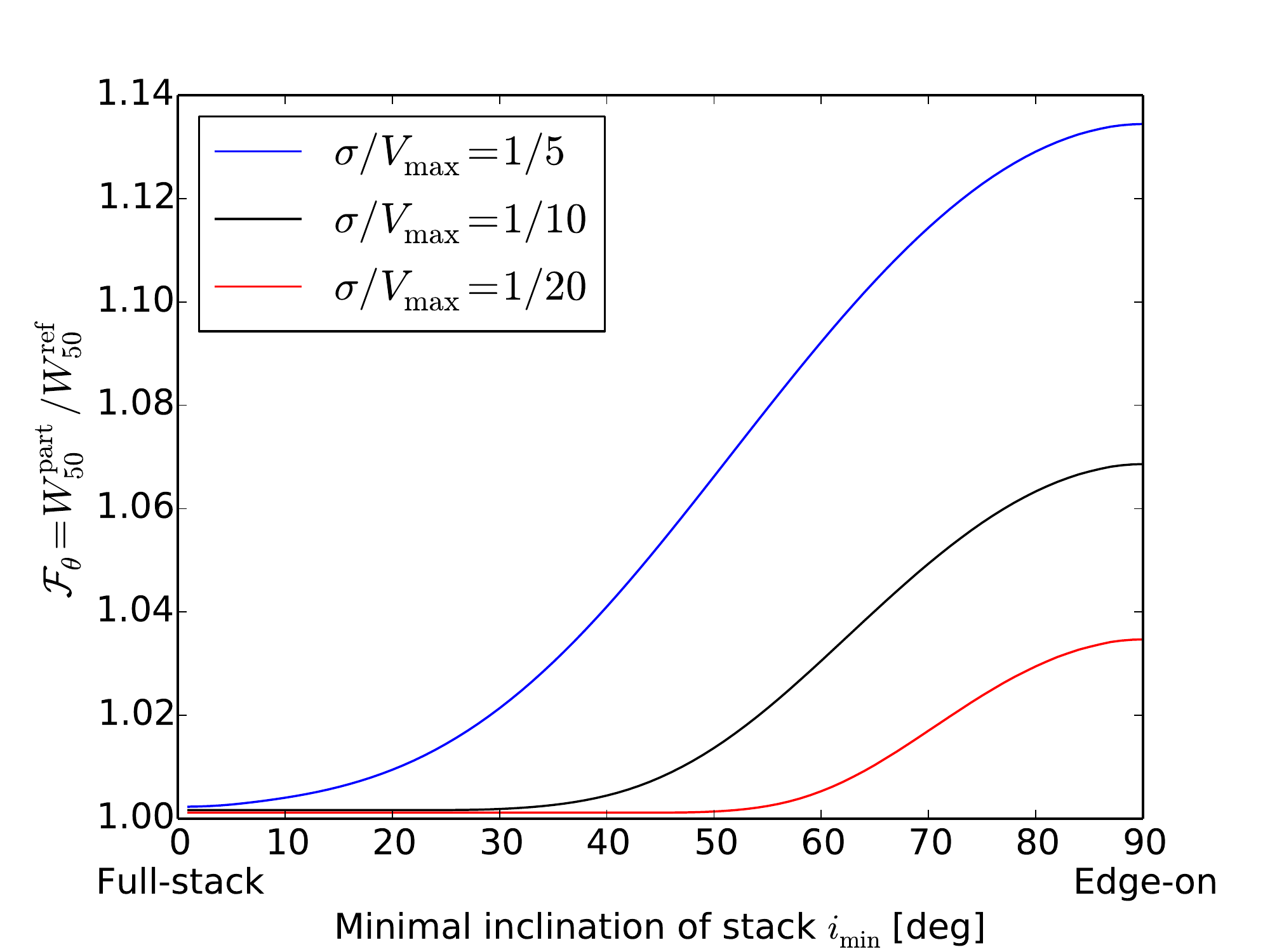}
	\caption{The width of stacked profiles normalised by deprojected and dispersed (left) and by deprojected and dedispersed (right) galaxy widths as a function of the minimum inclination of galaxy spectra included in the stack.\label{Plot A1b}}
	\par\end{centering}
\end{figure*}

The first case we investigate uses a simplistic galaxy model. This model serves as a first approximation to how the spectral widths of stacks relate to the width of the individual galaxies. We will also use this model to show the effect of gas dispersion on this relation.

We begin by calculating $\mathcal{F}$ using analytical emission profiles. Galaxies are modelled as disks with constant linear rotation velocities ({\it not} constant angular rotation velocities), implying normalised edge-on line profiles given by
\be\label{const rot}
	\rho_{\rm const}^{\rm edge}(v) = \frac{1}{\pi\sqrt{1-v^2}},
\ee
where $v = V/V_{\text{max}}$ is the normalised velocity and $V_{\rm max}$ is the linear velocity of the gas \citet{Stewart2014}. In this equation, $1/\sqrt{x}$ with $x<0$ is taken to be zero to avoid `if' conditions in the equation, and similarly in the equations that follow. This equation, and the following single spectrum equations, are normalised such that $\int_{-\infty}^{\infty}{\rm d}v~\rho=1~\forall~i$. To model an inclined galaxy, we substitute $V_{\text{max}}$ with $V_{\text{max}} \sin{i}$, where the $\sin{i}$ factor is due to the line-of-sight projection of the inclined disk. This produces
\be\label{inclined eqn}
	\rho_{\rm const}^{\rm incl}(v,i) = \frac{1}{\pi\sqrt{\sin^{2}{i}-v^2}}.
\ee
Note that this equation remains normalised, i.e. $\int \rho_{\rm const}^{\rm incl} = 1$. A lower $i$ leads to a more narrow but higher line. To simulate a stacked spectrum, we assumed a $\sin{i}$ inclination distribution, as expected for an isotropic, homogeneous universe. The equation describing a stack created in this way is given by
\be\label{stackeqn}
	\rho_{\rm const}^{\rm stack}(v) = \int_0^{\pi/2}\!\!\!\!{\rm d}i\, \frac{\sin i}{\pi\sqrt{\sin^2 i-v^2}}
\ee
The $\sin{i}$ factor in the numerator is a weight that ensures a $\sin{i}$ inclination distribution expected for random galaxy orientations in a 3D universe. It is interesting to note that equation~(\ref{stackeqn}) solves to a rectangular top-hat profile of value $\frac{1}{2}$ between $-1$ and $+1$, and value 0 otherwise (see Appendix) and thus $\int {\rm d}v ~ \rho_{\rm const}^{\rm stack}= 1$. Dispersed spectra are created by convolving $\rho(v)$ $(=\rho_{\rm const}^{\rm edge}(v)\text{ or }\rho_{\rm const}^{\rm stack}(v))$ with a Gaussian;
\be\label{add dispersion}
	\rho(v,s) = \int_{-\infty}^{\infty}{\rm d}v'~\frac{e^{-(v-v')^2/2s^2}}{s\sqrt{2\pi}}~\rho(v'),
\ee
where $s\equiv\sigma/V_{\text{max}}$ and $\sigma$ is the velocity dispersion of the gas. Because gas dispersion is assumed isotropic, it is immune to inclination effects and can always be added last, after changing the inclination of a profile or after creating a stacked profile.

To tackle in detail how a stacked profile builds up when successively adding galaxies of different inclinations, let us consider the partial stack
\be\label{part const}
	\rho_{\rm const}^{\rm part}(\theta,v) = \int_\theta^{\pi/2}\!\!\!\!{\rm d}i\, \frac{\sin i}{\pi\sqrt{\sin^2 i-v^2}},
\ee
where $\theta$ is the minimal inclination of the stack. Note that $\rho_{\rm const}^{\rm part}(\theta,v)$ is identical to $\rho_{\rm const}^{\rm stack}(v)$ if $\theta = 0$ and identical to $\rho_{\rm const}^{\rm edge}(v)$ if $\theta = \pi/2 = 90^\circ$. As with the other profiles, the partial stack can be dispersed via equation~(\ref{add dispersion}), giving the dispersed partial stack $\rho_{\rm const}^{\rm part}(\theta,v,s)$. Examples of model spectra $\rho_{\rm const}^{\rm edge}(v,s)$ (dashed lines), $\rho_{\rm const}^{\rm stack}(v,s)$ (solid lines) and $\rho_{\rm const}^{\rm part}(\theta = 45^\circ,v,s)$ (dash-dotted lines) can be seen in Fig.~\ref{Plot A1}. $\rho_{\rm const}^{\rm part}$ has an intermediate profile between the dispersed full stack (solid lines) and the dispersed edge-on profile (dashed lines). Equations~(\ref{const rot}), (\ref{stackeqn}), (\ref{add dispersion}) and (\ref{part const}) are all normalised (i.e. $\int \rho~{\rm d}i = 1$).

Given these expressions of line profiles, we can now explicitly identify $W_{50}^{\rm ref}$ and $W_{50}^{\rm stack}$ with the widths of $\mathrm{\rho_{const}^{edge}}(v)$ (without dispersion) and $\mathrm{\rho_{const}^{stack}(v,s)}$ (with dispersion s), respectively. Hence, $\mathcal{F}$ can be calculated analytically for any normalised dispersion $s$. By extension, we can calculate $\mathcal{F}_\theta = W_{50}^{\rm part}/W_{50}^{\rm ref}$, where $W_{50}^{\rm part}$ is the 50-percentile width of the dispersed partial stack with inclination limit $\theta$. Note that $\mathcal{F}_0 \equiv \mathcal{F}$.

Fig.~\ref{Plot A1b} shows $\mathcal{F}_\theta^{\text d}$ and $\mathcal{F}_\theta$ as a function of the minimal inclination $\theta$ for three different normalised dispersions $s=\sigma/V_{\rm max}$, covering the typical range. In fact, local galaxies typically have a $\sigma$ value of order 10 km s$^{-1}$ \citep{Ianjamasimanana2012} and an average $V_{\text{max}}$ of 50 - 200 km s$^{-1}$, hence a value of $s$ between $1/20-1/5$. The left panel shows $\mathcal{F}_{\theta}^{\text d} = W_{50}^{\text part}/W_{50}^{\text ref,d}$, where $W_{50}^{\text ref,d}$ is the deprojected spectral width of the galaxy {\it with dispersion}. This comparison suggests a correction dependent on $s$ is required to both dedisperse and correct for the stacking process.

The right panel of Fig.~\ref{Plot A1b}, however, shows $\mathcal{F}_{\theta} = W_{50}^{\text part}/W_{50}^{\text ref}$, where $W_{50}^{\text ref}$ is the deprojected {\it and dedispersed} line width for the galaxy. In this sense, we are bundling the dispersion correction and the stacking correction factor together. The remarkable result of this analysis is that $\mathcal{F}$ $(\equiv \mathcal{F}_0)$ is virtually identical to 1 for all realistic dispersions: {\it For galaxies with a fixed circular rotation velocity, the 50-percentile width of an isotropic stack of {\HI} lines is identical to the 50-percentile width of the input galaxies, if they were seen edge-on without dispersion. That is $W_{50}^{\rm stack} = 2V_{\rm max}$ in this model.}

This result suggests that the best way to measure $V_{\text{max}}$ is to stack galaxies of all inclinations. This allows us to measure $V_{\text{max}}$ without needing any inclination information and also gives us the same value regardless of the dispersion in the galaxies. The values of $s$ used in Fig.~\ref{Plot A1b} should cover most realistic cases in the local Universe, but even with a dispersion as large as $s = 1/2$, $\mathcal{F}$ still equals $\sim1.04$.

This investigation using disk galaxies with constant circular rotation velocities shows the stacking technique has merit in reproducing the spectral widths needed to construct the TFR. We now need to investigate more realistic galaxies to see if any corrections need to be made and track where those corrections come from.

\subsection[]{Disk galaxy with varying circular rotation velocity}
\label{galaxies with rotation curves}

\begin{figure*}
	\begin{centering}
	\includegraphics[scale=0.45]{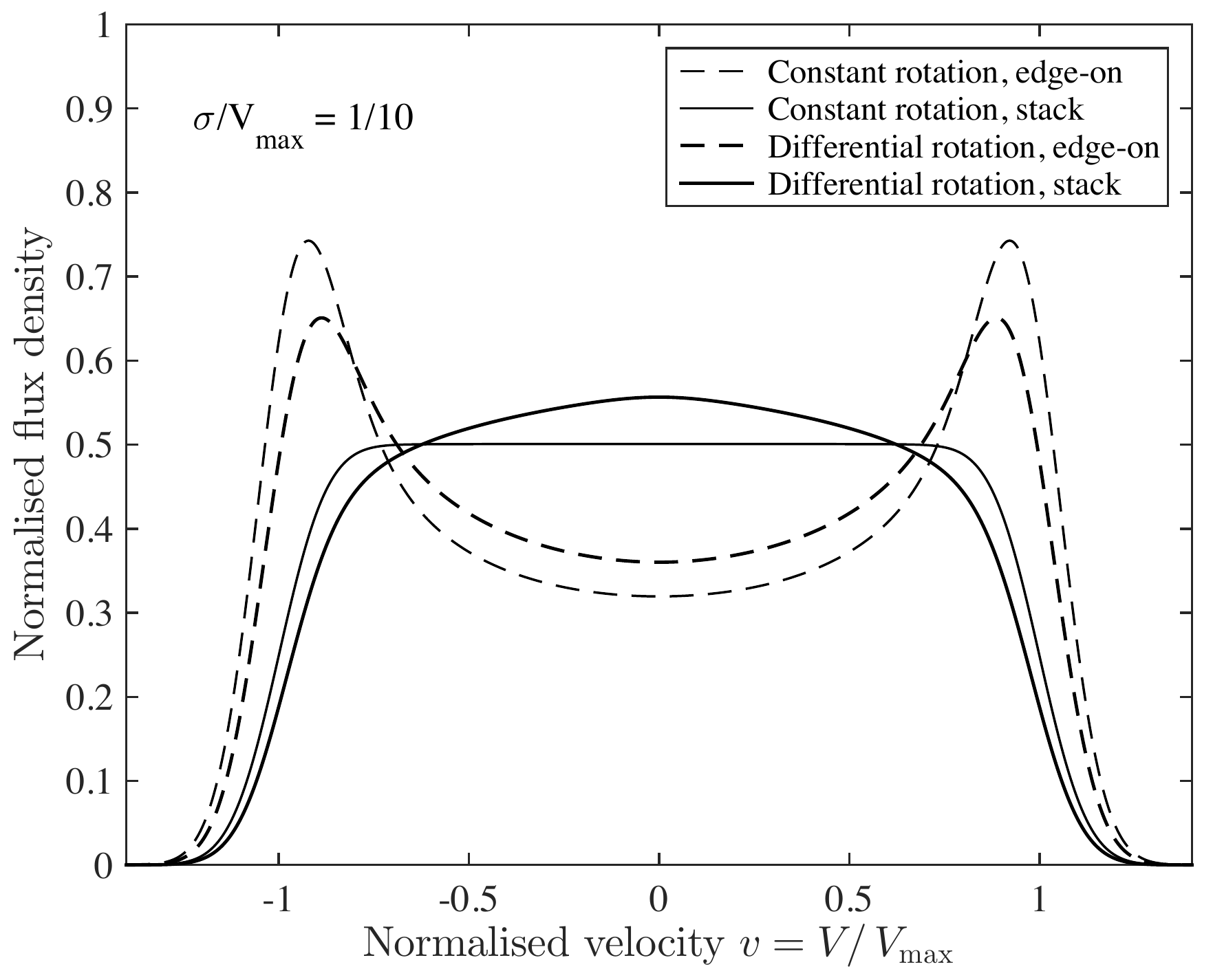}\hspace{1cm}
	\includegraphics[scale=0.45]{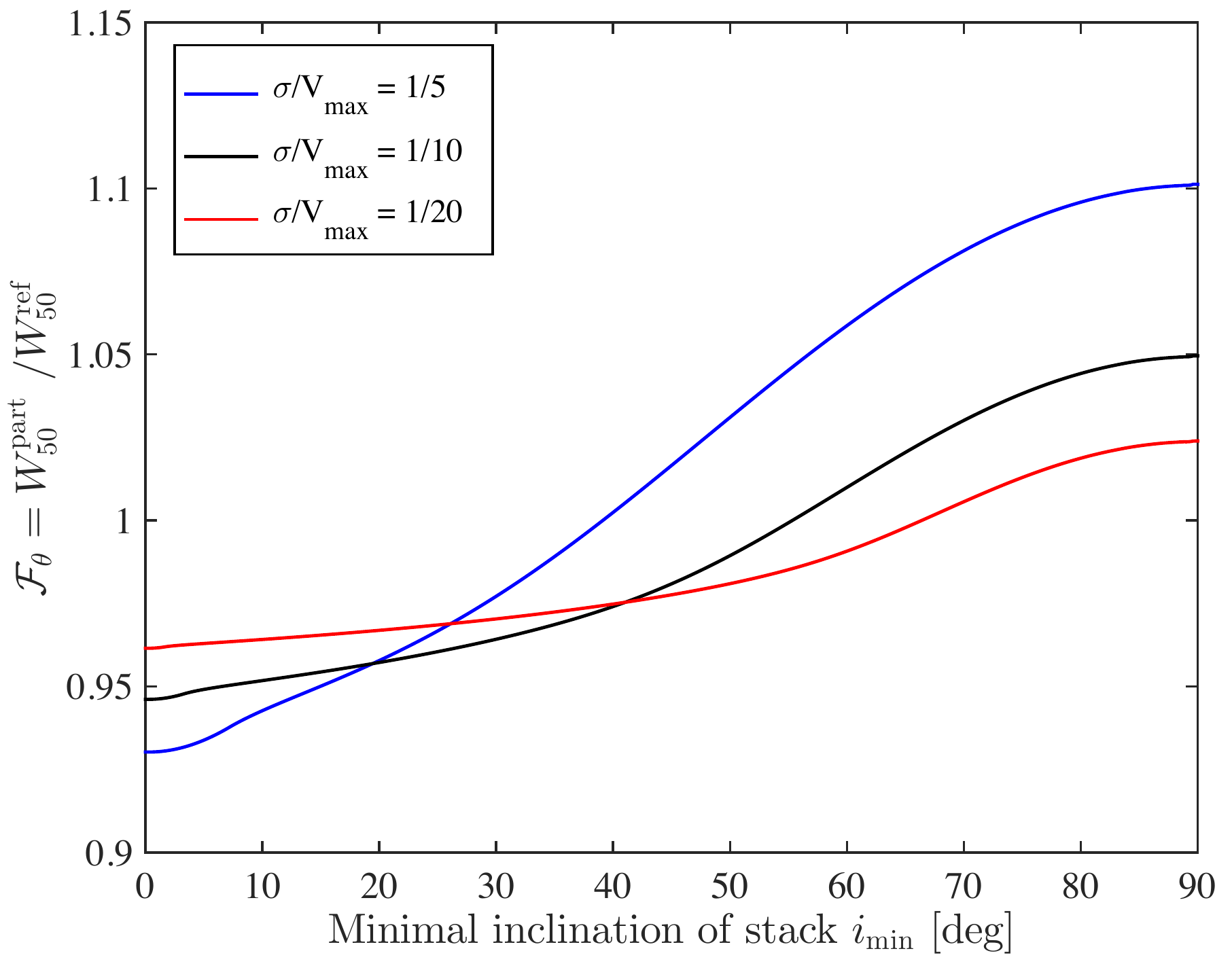}
	\caption{\label{Plot A2}(Left) the emission profile of an edge-on galaxy created with an exponential circular velocity rotation curve and gas mass profile using equations~(\ref{diff rotation}) and (\ref{add dispersion}) shown with a thick dashed line. The thick solid line is the stack of these emission lines as observed at all possible inclinations weighted by $\sin{i}$ to mimic the inclination distribution expected from an isotropic homogeneous universe. This emission profile line is created by combining equations~(\ref{diff rotation stack}) and (\ref{add dispersion}). Shown in thin lines is the emission profile of a single galaxy and a stack of galaxies from \S\ref{constant rotation velocity galaxies} using equations~(\ref{const rot}), (\ref{stackeqn}) and (\ref{add dispersion}). All profiles were created using a normalised dispersion value of $\sigma/V_{\text max} = 1/10$. (Right) the width of stacked profiles normalised by dispersion-less edge-on galaxy width as a function of the minimum inclination of the galaxies included in the stack. All stacks were created using equations~(\ref{diff rotation stack}) and (\ref{add dispersion}).}
	\par\end{centering}
\end{figure*}

Assuming all concentric gas rings for the disks described in \S\ref{constant rotation velocity galaxies} have the same rotation velocity does not accurately describe the inner parts of realistic rotation curves. In this section we analyse more realistic galaxy models based on a differential rotation curve of the form
\begin{equation}
	V(r) = V_{\text{max}} \left(1 - e^{-r/r_{\rm flat}}\right),
\end{equation}
where r is the galactocentric radius, $r_{\rm flat}$ is the characteristic scale length of the rotation curve and $V_{\rm max}$ is the asymptotic velocity as $r\gg r_{\rm flat}$.

Given differential rotation, the {\HI} gas is subjected to different Doppler shifts, depending on where in the rotation curve this gas lies. Therefore, we now have to specify the surface density profile $\Sigma_{\rm HI}(r)$ of the {\HI} gas, unlike in the case of a constantly rotating disk (\S\ref{constant rotation velocity galaxies}). We therefore adopt the standard model of an exponential disk;
\begin{equation}
	\Sigma_{\rm HI}(r) = \frac{M_{\rm HI}}{2 \pi r_{\rm HI}^2}e^{-r/r_{\rm HI}},
\end{equation}
where $r_{\rm HI}$ denotes the characteristic scale length of the {\HI} disk. A value of $r_{\rm HI}/r_{\rm flat} = 3$ is used, corresponding to the average value from the THINGS catalogue \citep{Leroy2008}.

Adding these two physical profiles to the equations derived in \S\ref{constant rotation velocity galaxies} we get the following equation describing the emission profile of a single galaxy;
\be\label{diff rotation}
	\rho_{\rm diff}^{\rm edge}(v) = \int_0^{\infty}\!\!\!{\rm d}r\, \frac{r\,e^{-r}}{\pi\sqrt{[1-e^{-3r}]^2-v^2}},
\ee\vspace{2mm}
and for a stack of these galaxies;
\be\label{diff rotation stack}
	\rho_{\rm diff}^{\rm stack}(v) = \int_0^{\pi/2}\!\!\!\!{\rm d}i~\int_0^{\infty}\!\!\!{\rm d}r\, \frac{r\,e^{-r}\,\sin i}{\pi\sqrt{[1-e^{-3r}]^2\sin^2 i-v^2}}
\ee
assuming a $\sin{i}$ inclination distribution and normalised velocity $v$ ($= V/V_{\text{max}}$). The `3' coefficient in the denominator's exponent comes from the ratio of the mass scale radii and velocity ($r_{\rm HI}/r_{\rm flat}$). To produce the corresponding $\rho_{\rm diff}^{\rm edge}$ and $\rho_{\rm diff}^{\rm stack}$ lines with normalised Gaussian velocity dispersion $s$ ($= \sigma/V_{\text{max}}$), we use equation~(\ref{add dispersion}).

Fig.~\ref{Plot A2} (left) shows a single galaxy spectrum and a stacked spectrum using equations~(\ref{const rot}), (\ref{stackeqn}) and (\ref{add dispersion}) (disks with constant rotation), as well as a single galaxy profile and a stacked spectrum using equations~(\ref{diff rotation}), (\ref{diff rotation stack}) and (\ref{add dispersion}) (disks with varying differential rotation).

In order to calculate $\mathcal{F}_\theta$ ($= W_{50}^{\rm part}/W_{50}^{\rm ref}$) as we did in \S\ref{constant rotation velocity galaxies}, we define the partial stack function as
\be\label{part diff}
	\rho_{\rm diff}^{\rm part}(\theta,v) = \int_\theta^{\pi/2}\!\!\!\!{\rm d}i~\int_0^{\infty}\!\!\!{\rm d}r\, \frac{r\,e^{-r}\,\sin i}{\pi\sqrt{[1-e^{-3r}]^2\sin^2 i-v^2}},
\ee
where $\theta$ is the minimal inclination of the stack. Analogous to equation~(\ref{part const}), $\rho_{\rm diff}^{\rm part}(0,v) \equiv \rho_{\rm diff}^{\rm stack}$ and $\rho_{\rm diff}^{\rm part}(\pi/2,v) \equiv \rho_{\rm diff}^{\rm edge}$. The partial stack can be dispersed using equation~(\ref{add dispersion}) giving the dispersed partial stack $\rho_{\rm diff}^{\rm part}(\theta,v,s)$. Equations~(\ref{diff rotation}), (\ref{diff rotation stack}) and (\ref{part diff}) are normalised (i.e. $\int \rho~{\rm d}i = 1$).

In Fig.~\ref{Plot A2} (right) we show $\mathcal{F}_\theta$ as a function of minimal inclination $\theta$ for a range of different normalised dispersions $s = \sigma/V_{\rm max}$. This produces a value of $\mathcal{F} = 0.95 \pm 0.01$. The errors stated in the value of $\mathcal{F}$ come from the variation with $s$. This value is slightly smaller than $\mathcal{F} = 1$ for galaxies with constant rotation in \S\ref{constant rotation velocity galaxies}.

We have now established that we expect a 5\% offset in $\mathcal{F}$ due to the fact that, when stacked, {\HI} spectra with realistic rotation and mass profiles do not produce perfect rectangular top-hats. We now investigate the effect low-number statistics play in stacking.

\subsection[]{Low-number effects}
\label{low-number effects}

\begin{figure*}
	\begin{centering}
	\includegraphics[scale=0.43]{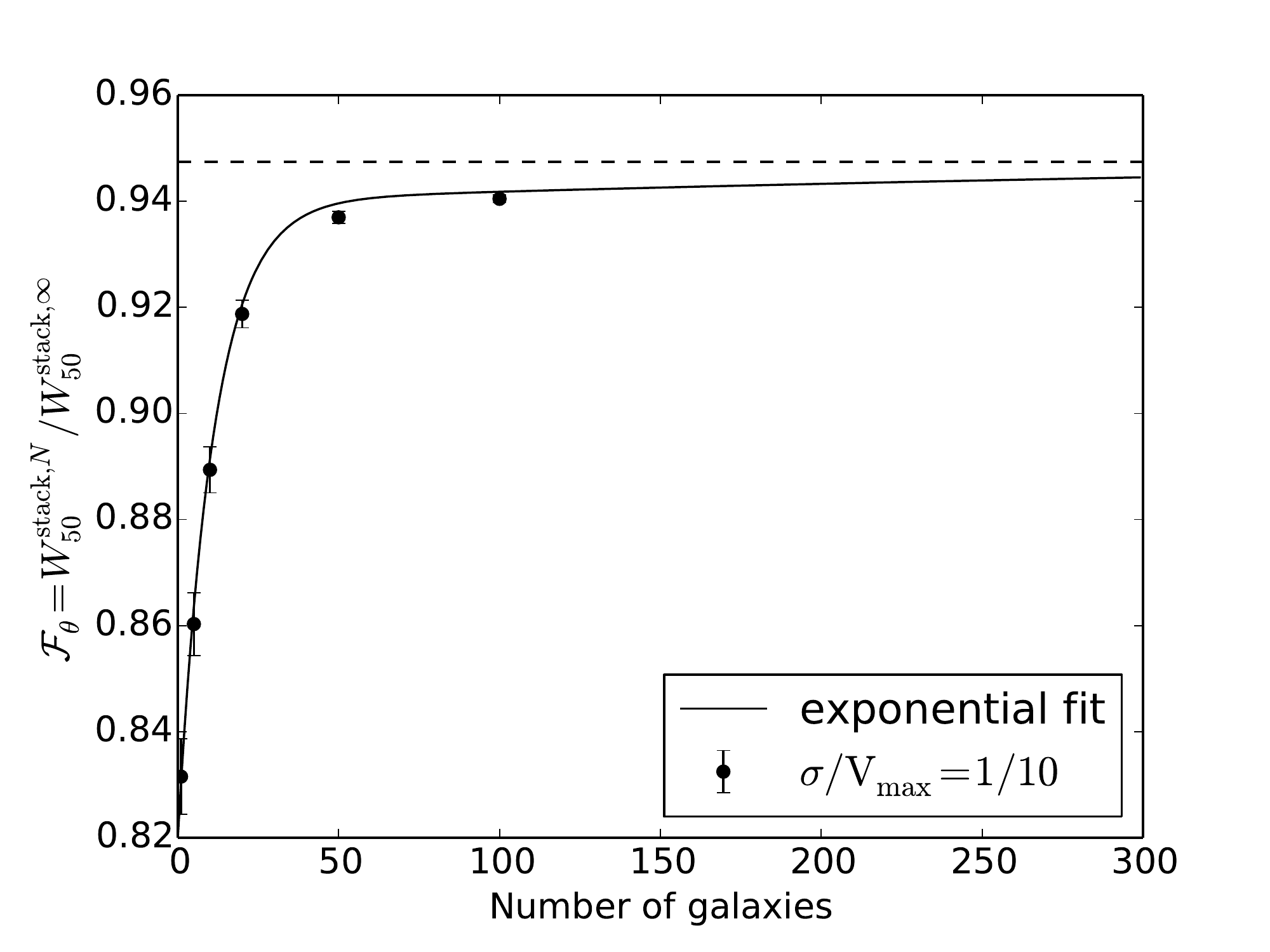}
	\includegraphics[scale=0.43]{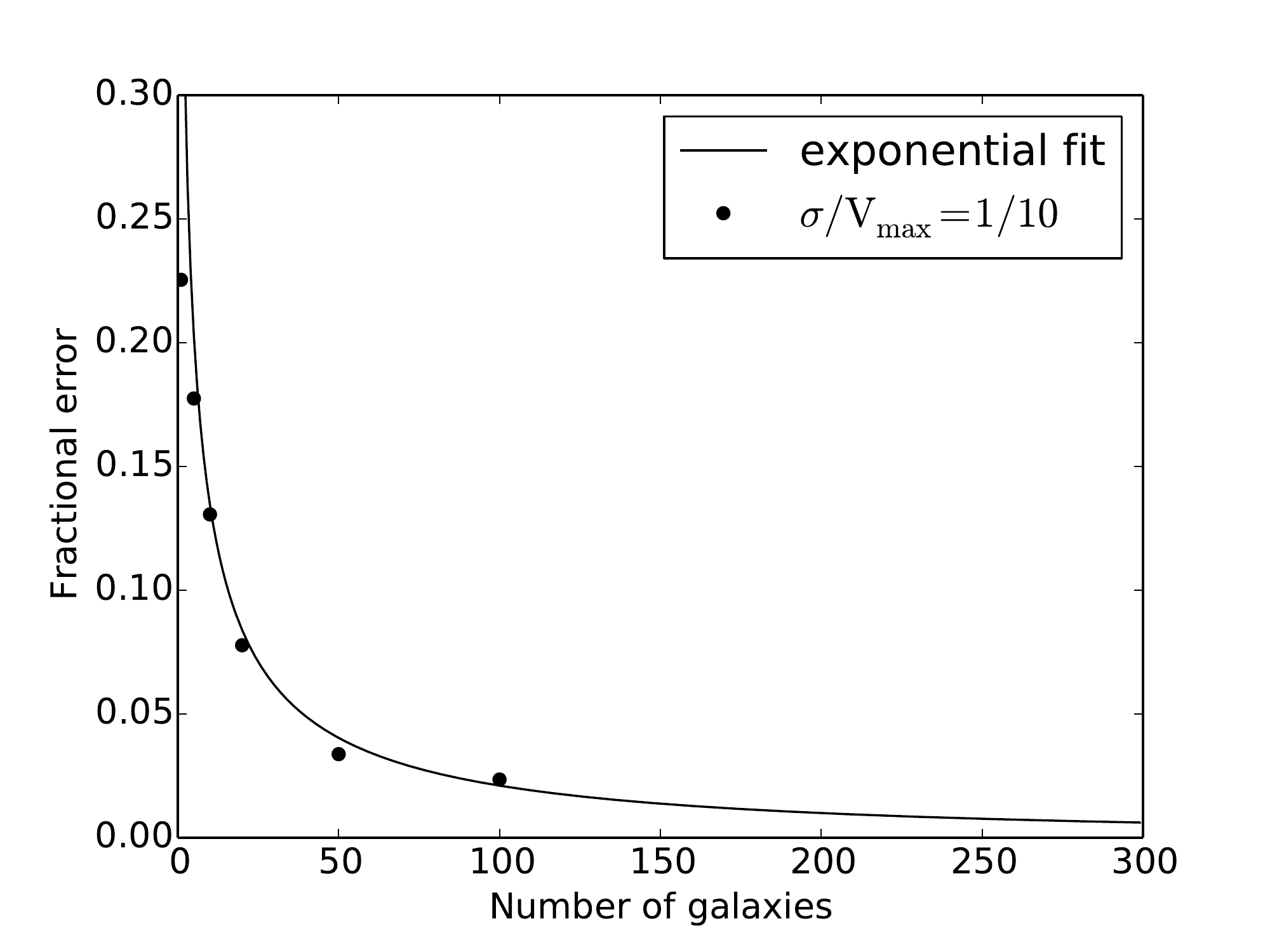}
	\caption{ \label{Plot A3}Using simulated galaxies like those described in \S\ref{galaxies with rotation curves}, the above two frames each show a different type of offset in the spectral stacks as a function of how many galaxies are included in the stack. (Left) the correction factor $\mathcal{F}$, which is a measure of the systematic offset of $W_{50}^{\rm stack,{\it N}}$ compared to $2V_{\text{max}}$, was measured for stacks of 1, 5, 10, 20, 50, 100 and infinitely many galaxies. These measurements were repeated one-thousand times and the mean offset is shown as black data points. The asymptotic value, represented as a dashed line, is $\mathcal{F}$ for a stack of infinitely many galaxies. These data were fit using equation~(\ref{exp 1}). (Right) the 1 $\sigma$ standard deviation of $W_{50}^{\rm stack,{\it N}}$ from $W_{50}^{\rm stack,\infty}$ as a function of the number of galaxies per stack. We fit equation~(\ref{exp 2}) to the data.}
	\par\end{centering}
\end{figure*}

In reality, not every single stack contains a large number of galaxies, causing deviations from the pure $\sin{i}$ inclination distribution assumed in \S\ref{constant rotation velocity galaxies} and \S\ref{galaxies with rotation curves}. Low number statistics may additionally cause the final stack to be dominated by a few galaxies with large fluxes. In this section we assess the impact of these effects on the widths of stacked {\HI} profiles.

We ran simulations to measure the statistical effects on $\mathcal{F}$ as a function of $N$, the number of galaxies included in the stack, using the galaxies described in \S\ref{galaxies with rotation curves} (equations~\ref{diff rotation}, \ref{diff rotation stack} and \ref{add dispersion}). In these simulations, $N$ galaxies were selected at random from a $\sin{i}$ inclination distribution. The galaxies were stacked and the width of the spectrum created ($W_{50}^{\rm stack,{\it N}}$) was compared to the width of the spectrum created using infinite galaxies ($W_{50}^{\text{stack,}\infty}$) and $W_{50}^{\rm ref}$ to quantify the errors and $\mathcal{F}$ respectively. The $N$ galaxies were re-picked and stacked 1000 times to gain a statistically significant sample. This process was repeated with $N = 1, 10, 20, 50$ and $100$ galaxies.

We found a systematic and random error component to $\mathcal{F}$ which are both a function of $N$. Fig.~\ref{Plot A3} shows the data from 1000 samples and the functional fit to both the systematic component (left) and the random component (right). The systematic shift in $\mathcal{F}$ occurs as no combination of galaxies will create a stack where $W_{50}^{{\rm stack,}N} > W_{50}^{\rm stack,\infty}$, but $W_{50}^{{\rm stack,}N}$ can be lower than $W_{50}^{\rm stack,\infty}$ if edge-on galaxies are missing from the sample.

Both the systematic and random errors become smaller with more galaxies, as expected, and both approach reasonable values as the number of galaxies increases. The systematic offset approaches the value for infinite galaxies found in \S\ref{galaxies with rotation curves} $(\approx 0.95$ for $s = 1/10)$ and the Gaussian scatter in widths approaches 0. We fit exponential functions;
\begin{equation}
\label{exp 1}
	\mathcal{F}=0.95  -0.12~e^{-0.08~N} -0.01~e^{-0.003~N}
\end{equation}
and
\begin{equation}
\label{exp 2}
	\mathrm{error}=3.2~W_{50}^{\rm stack} ~e^{-2.0~N^{0.2}} ~ V_{\rm max}^{-1},
\end{equation}
to the offset and error relations respectively.

The corrections presented in this section are not used in \S\ref{Millennium simulated galaxies} as each stack contains several thousand galaxies, so the corrections are negligible. \S\ref{sensitivity-limited simulated galaxies} onwards do, however, use equation~(\ref{exp 1}) and include equation~(\ref{exp 2}) in the error calculations.

\section[]{Simulated galaxies}
\label{Simulated galaxies}

In this section we consider the line widths $W_{50}^{\rm stack}$ of stacked emission lines composed of a {\it distribution of different model galaxies} that better represent the diversity of galaxies in the Universe, rather than simply stacking a single randomly oriented {\HI} profile. To study $\mathcal{F}$ (equation~\ref{eq F}) we need to define the line width $W_{50}^{\rm ref}$. As we are no longer using identical galaxies, $W_{50}^{\rm ref}$ is now defined as the mass weighted geometric average over the dedispersed edge-on line widths of the galaxies used in the stack. In the next two subsections we study $\mathcal{F}$, using a volume (\S\ref{Millennium simulated galaxies}) and sensitivity (\S\ref{sensitivity-limited simulated galaxies}) limited mock sample. We then add noise to our galaxies in \S\ref{Gaussian noise} and investigate how much noise our technique can cope with.

Both of these samples were generated from the S$^3$-SAX model \citet{Obreschkow2009b}, which is a semi-analytic model (SAM) for {\HI} and ${\rm H_2}$ in galaxies. This model builds on the SAM by \citep{DeLucia2007}, which uses formulae based on empirical or theoretical considerations to simulate gas cooling, re-ionisation, star formation, supernovae (and associated gas heating), starbursts, black holes (accretion and coalescence), and the formation of stellar bulges due to disk instabilities.

\subsection[]{Volume-complete simulated sample}
\label{Millennium simulated galaxies}

\begin{figure*}
\begin{centering}
\includegraphics[scale=0.43]{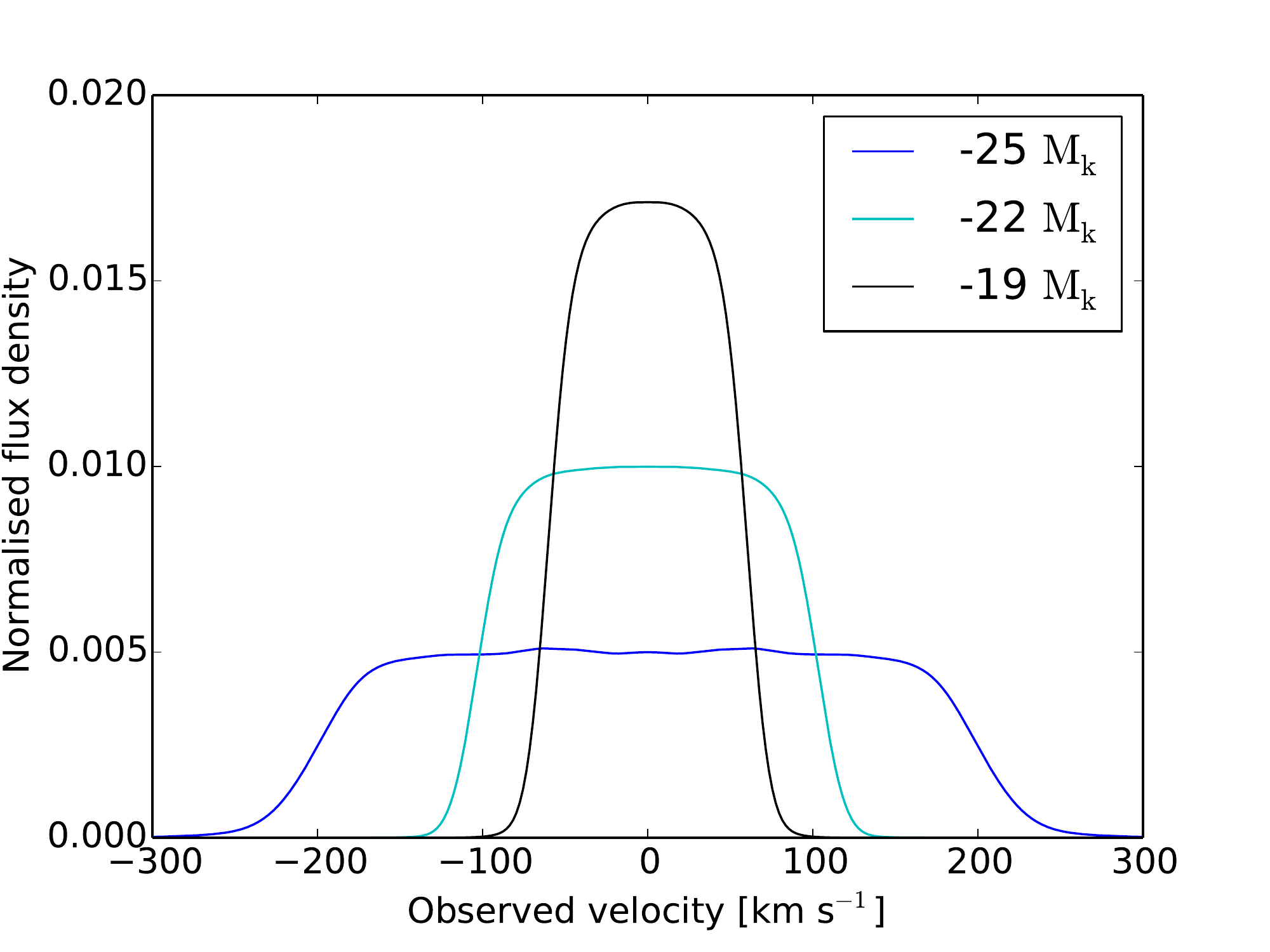}
\includegraphics[scale=0.43]{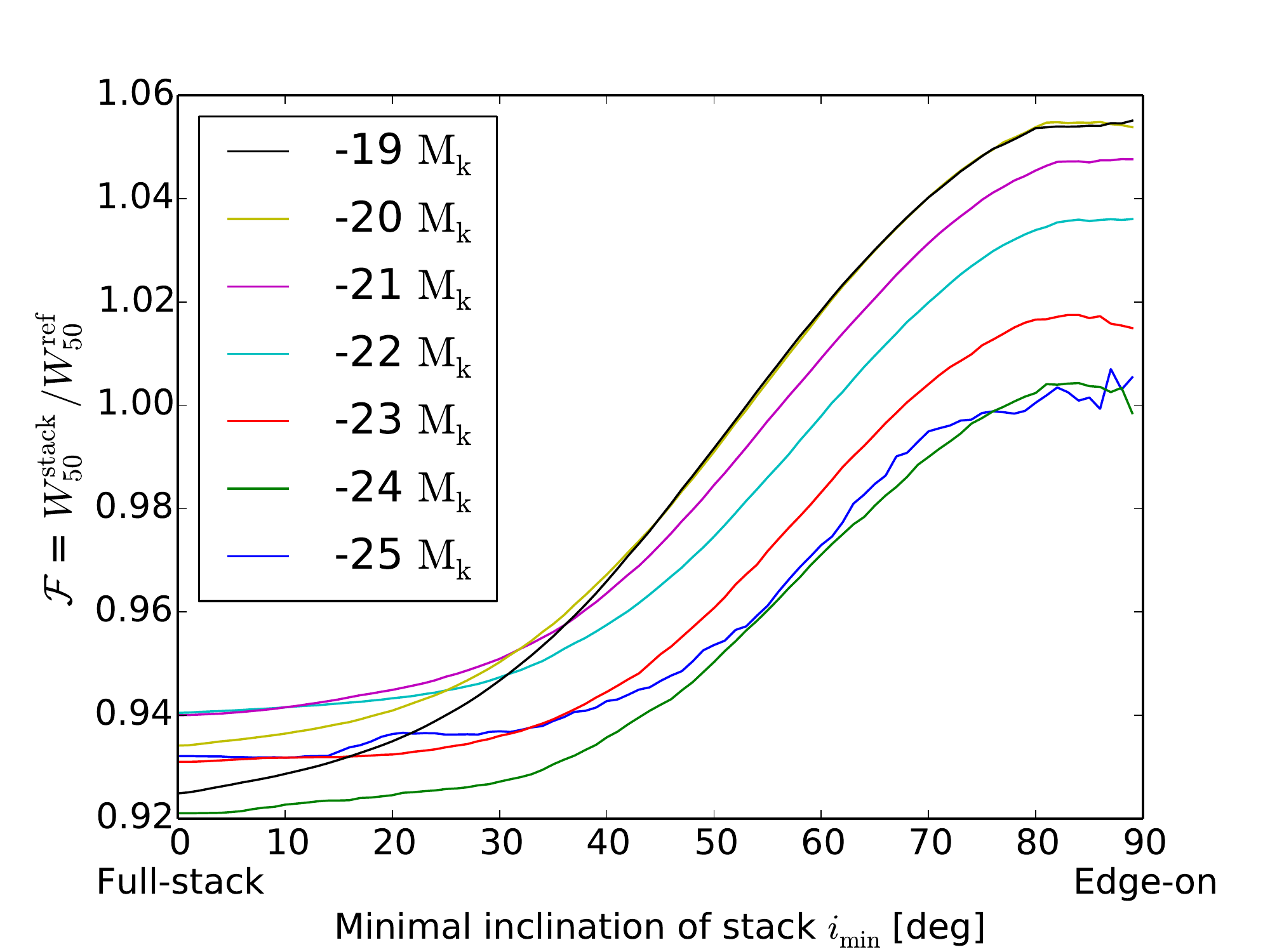}
\caption{(Left) three normalised stacked spectra. The two most extreme magnitude bins are shown as well as the middle magnitude bin. (Right) the width of the stacked spectra, normalised to the width of corresponding edge-on spectra, as a function of the minimum inclination included in the stack. Every galaxy in the S$^3$-SAX is included, with the exception of galaxies whose morphology is known to be elliptical or irregular.\label{Plot B1}}
\par\end{centering}
\end{figure*}

\begin{figure*}
\begin{centering}
\includegraphics[scale=0.45]{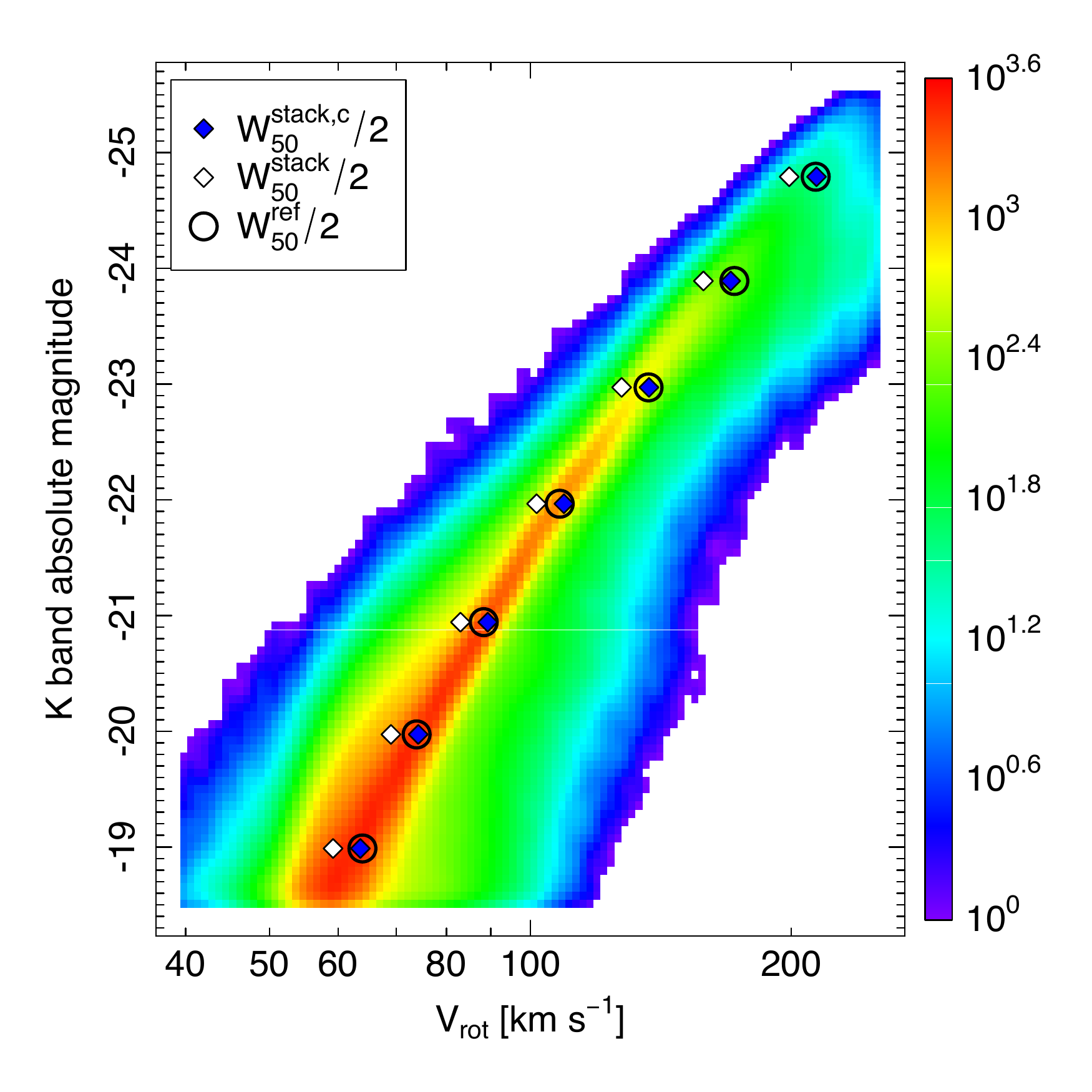}
\includegraphics[scale=0.45]{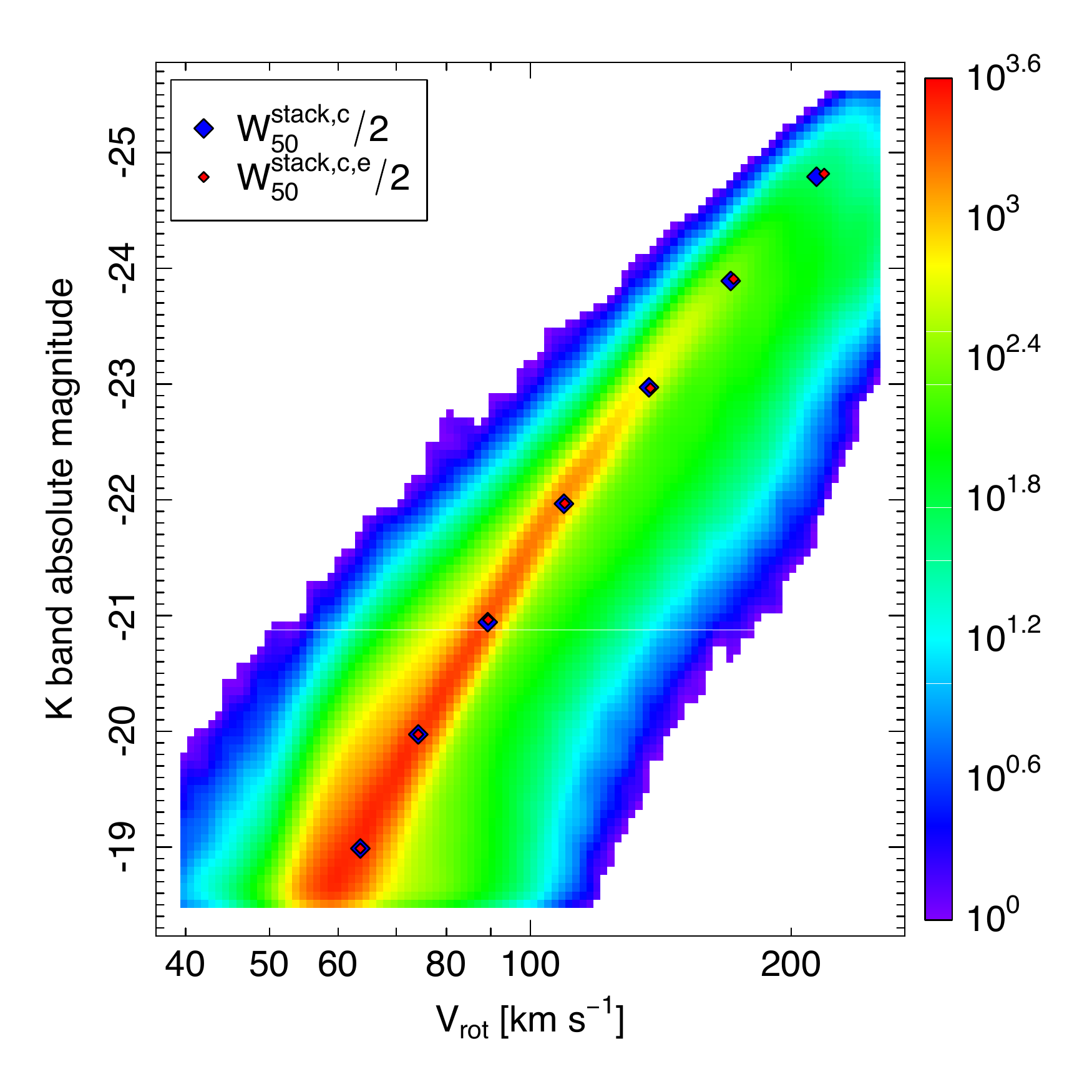}
\caption{The Tully-Fisher relation using S$^3$-SAX galaxies based on the Millennium simulation. Individual galaxies are represented as a log density map. Excluding resolved non-spiral galaxies (left), the white diamonds are uncorrected $W_{50}/2$ values measured from each stack. The blue diamonds are $W_{50}/2$ values measured from each stack after being corrected by $\mathcal{F}^{-1} = 0.93^{-1}$. The large black circles indicate the geometric mean $W_{50}/2$ measured from galaxies that have been corrected for inclination and dedispersed. Including resolved non-spiral galaxies (right), the large blue diamonds are $W_{50}/2$ values measured from each stack after excluding known ellipticals and being corrected by $\mathcal{F}^{-1} = 0.93^{-1}$. The smaller red diamonds are $W_{50}/2$ values measured from each stack including elliptical galaxies after being corrected by $\mathcal{F}^{-1} = 0.93^{-1}$. In both plots the error bars are omitted as they are smaller than the data points.\label{Plot B2}}
\par\end{centering}
\end{figure*}

Mass spectra were produced using raw output from the spiral galaxies in the S$^3$-SAX before dispersion effects were added. Each galaxy was then assigned a random viewing inclination. No noise was added to these spectra. Spectra were binned into seven equally spaced magnitude bins with a width of 1 magnitude. These dispersion-less spectral profiles were deprojected, and then had their widths measured. These widths were weighted by the {\HI} mass of the corresponding galaxy producing the spectrum, and finally geometrically averaged to produce $W_{50}^{\rm ref}$.

Galaxy selection was kept to a minimum to prove the stacked TFR holds for all types of spiral galaxies with a complete sample of realistic galaxies. The resolution limit of the Millennium simulation \citep{Springel2005} is $8.6\times10^8~M_\odot$, which sets the completeness limit for the S$^3$-SAX to $M_{\rm HI}+M_{\rm H_2}\approx 10^8~M_\odot$ \citep{Obreschkow2009a}. Galaxies identified as ellipticals were excluded, however, only the brighter galaxies (absolute magnitudes less than approximately $-21$) could be morphologically classified, so there were still elliptical galaxies remaining in the data set. Elliptical galaxies do not make a significant impact in the resulting stacks due to their lower {\HI} content.

The emission profile of each galaxy was scaled down in frequency space by a factor of $\sin{i}$, and the flux density was scaled up by the same factor, giving each galaxy an inclination dependant spectral profile. These profiles were then smoothed with a Gaussian corresponding to a dispersion of $\sigma = 8 ~\text{km}~\text{s}^{-1}$, just as in \citet{Obreschkow2009b}. After being sorted into magnitude bins, galaxies with inclinations between $\theta$ and 90$^\circ$ were stacked to create partial stacks. The width of each partial stack is called $W_{50}^{\rm part}$ and $W_{50}^{\rm ref}$ is defined as the geometric average of the dedispersed edge-on {\HI} line widths of the subsample of galaxies included in the partial stack. Fig.~\ref{Plot B1} (left) shows examples of the stacked spectra produced, while Fig.~\ref{Plot B1} (right) shows $\mathcal{F}_\theta$ ($=W_{50}^{\rm part}/W_{50}^{\rm ref}$) as a function of $\theta$.

To create a TFR from our stacked spectra, we need to correct $W_{50}^{\rm stack}$ by the correction factor we have found. We do this by multiplying $W_{50}^{\rm stack}$ by $\mathcal{F}^{-1}$. We add a $c$ superscript ($W_{50}^{\rm stack,c}$ in this case) to denote a width corrected in this way. Due to only small differences in $\mathcal{F}$ for different magnitude bins, and the lack of any systematic trend, a global value for the correction factor was used. This value, which can be read from Fig.~\ref{Plot B1} (right), is $\mathcal{F} = 0.93\pm0.01$, which approximately agrees with the value of $0.95\pm0.01$ found in \S\ref{galaxies with rotation curves} using a simplistic analytical galaxy model. Hence, our calibrated width $W_{50}^{\rm stack,c}$ is given by $\mathcal{F}^{-1}W_{50}^{\rm stack} = W_{50}^{\rm stack}/0.93$.

Fig.~\ref{Plot B2} (left) is a density map consisting of individual galaxies from the S$^3$-SAX simulation. The position of an individual galaxy on the TFR is determined by its K-band absolute magnitude, and its dispersion-less edge-on $W_{50}$. Individual galaxies are represented by a density map while the corrected ($V_{\rm rot} = ~\mathcal{F}^{-1}W_{50}^{\rm stack}/2,~\mathcal{F} = 0.93$) and uncorrected ($V_{\rm rot} = W_{50}^{\rm stack}/2$) stack widths are displayed in blue and white diamonds respectively. Both stacked data sets use the average K-band absolute magnitude of the individual galaxies as their K-band absolute magnitudes. The black circles give an idea of where our values of $W_{50}^{\rm ref}$ lie on a TFR with respect to the underlying galaxy population they are derived from. The corrected values $W_{50}^{\rm stack,c}$, show very good agreement with both $W_{50}^{\rm ref}$ and the underlying galaxy distribution.

A second simulated subset was used which did not include an elliptical galaxy cut, shown in Fig.~\ref{Plot B2} (right). The inclusion or exclusion of the known elliptical galaxies in this data set makes little ($<1\%$) difference in the width of stacked data points, despite accounting for up to 62.7\% of the galaxies in some stacks. This is easily seen by comparing the small red diamonds (stack widths that include elliptical galaxies) to the blue diamonds (which are identical to those from Fig.~\ref{Plot B2}, left).

\subsection[]{Sensitivity-limited simulated sample}
\label{sensitivity-limited simulated galaxies}

\begin{figure}
\begin{centering}
\includegraphics[scale=0.45]{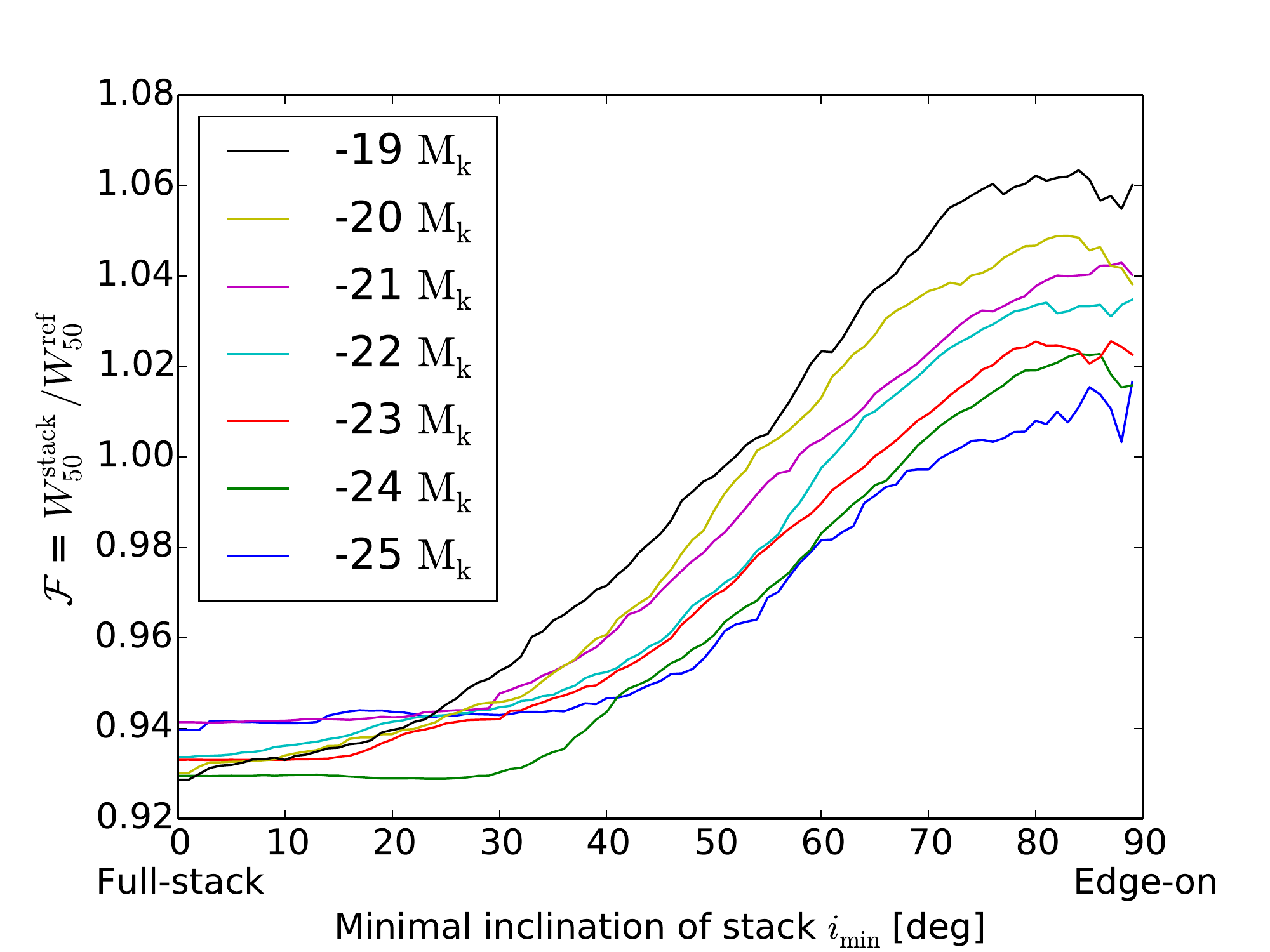}
\caption{The stacked profile widths normalised by dispersion-less edge-on galaxy widths as a function of the minimum inclination included in the stack for each magnitude bin in the HIPASS simulation.\label{Plot B5}}
\par\end{centering}
\end{figure}

\begin{figure}
\begin{centering}
\includegraphics[scale=0.45]{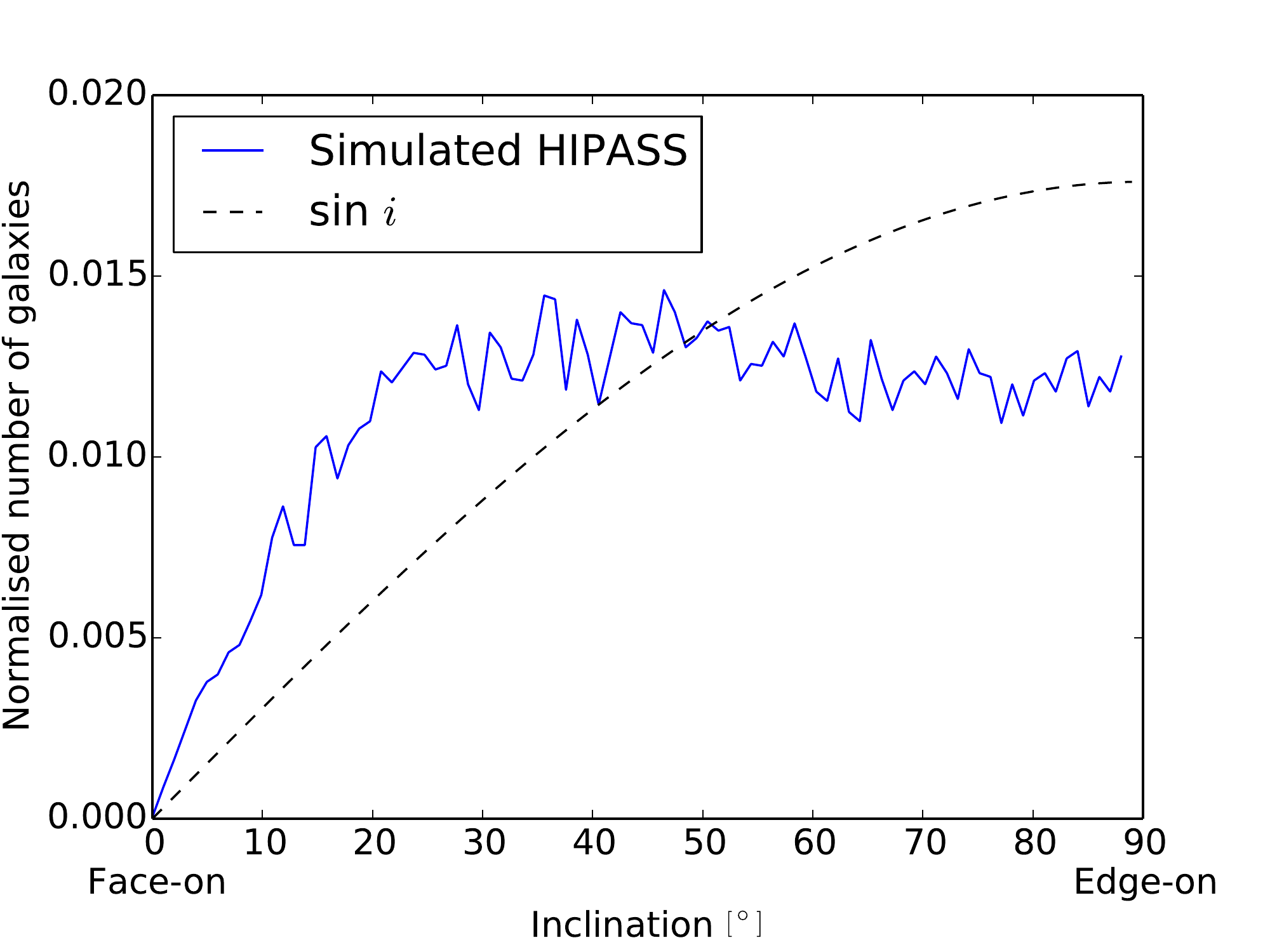}
\caption{The inclination distribution functions for the S$^3$-SAX galaxies: The dashed line is the inclination distribution for the complete Millennium box which follows a $\sin{i}$ inclination distribution (the volume-complete sample used in \S\ref{Millennium simulated galaxies}). The solid blue line is the inclination distribution for the simulated HIPASS samples (the sensitivity-limited sample used in \S\ref{sensitivity-limited simulated galaxies}).\label{Plot B6}}
\par\end{centering}
\end{figure}

\begin{figure}
\begin{centering}
\includegraphics[scale=0.45]{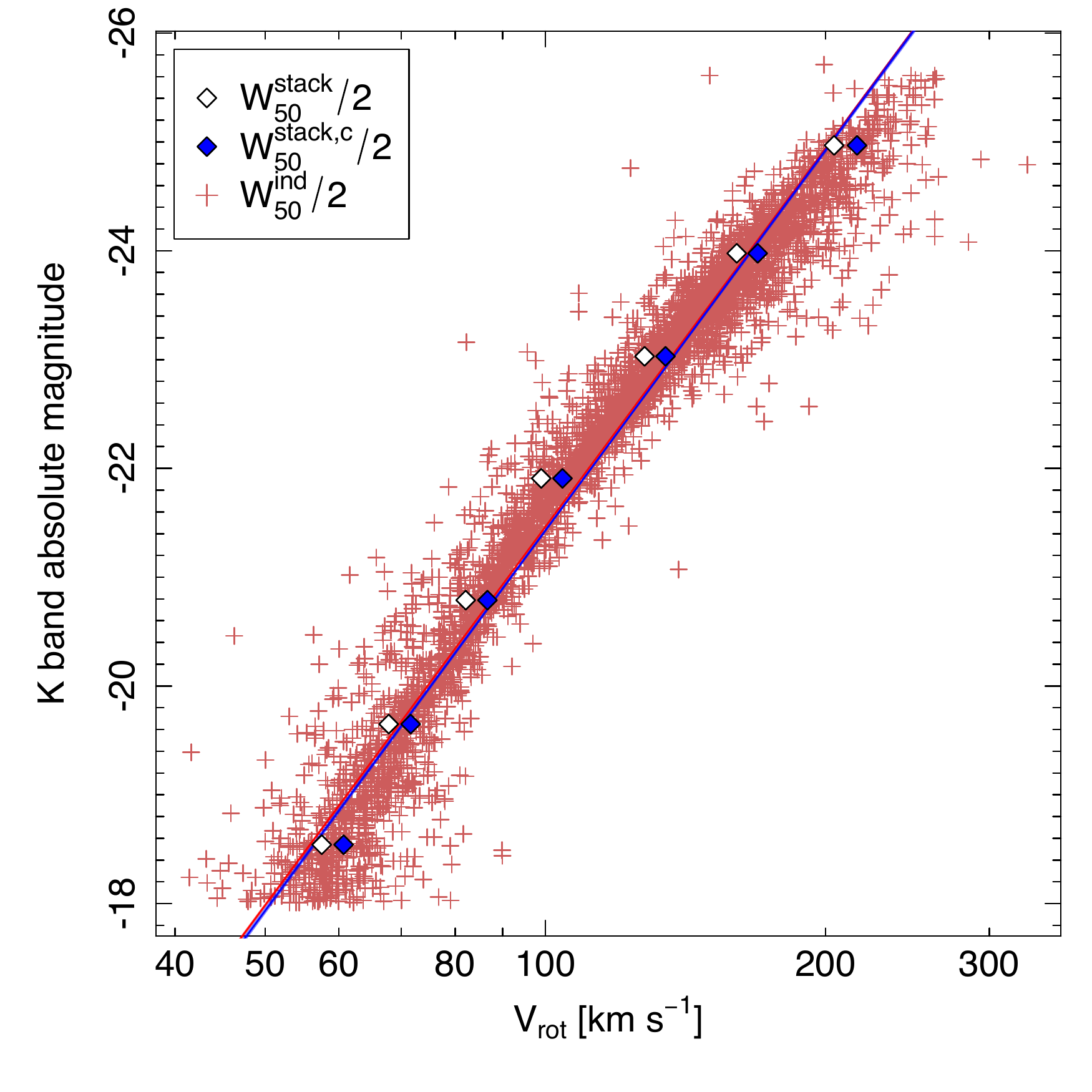}
\begin{tabular}{c c c c} 
\hline 
Relation & Slope & Offset \\ [0.5ex] 
\hline 
Stacked & $-11.61 \pm 0.04$ & $1.78 \pm 0.09$ \\
Individual & $-11.55 \pm 0.04$ & $1.64 \pm 0.09$ \\ [1ex]
\hline\hline 
\end{tabular}
\caption{The Tully-Fisher relation for a simulated HIPASS created using the S$^3$-SAX galaxies. Each of the 4260 red data points represents a galaxy on the Tully-Fisher plane with a rotation velocity given as the half-width at half-height of the emission spectra without dispersion effects. The white diamonds are the half-width at half-height of the stacked spectra while the blue diamonds have been corrected by the factor given in equation~(\ref{exp 1}) which for most stacks was $\mathcal{F}^{-1} = 0.95^{-1}$. Error bars are omitted as they are smaller than the data points. The red line is the average fit to the individual galaxies (red crosses) across all five HIPASS simulations, while the blue line is the average fit to the stacked data points (blue diamonds) across all five simulations. The slope and offset values listed are of the form used in equation~(\ref{TFRslope}).\label{Plot B4}}
\par\end{centering}
\end{figure}

We next investigate an {\HI} sensitivity-limited observational subset of the S$^3$-SAX, allowing us to make a comparison with the HIPASS data. To this end, five simulations were created using the HIPASS selection function.

To create these five simulations, the cubic simulation box of the Millennium simulation was divided into five, non-overlapping HIPASS-like volumes, as explained in \citet[\S2.2 \& Fig.~3]{Obreschkow2013a} containing between 3475 and 4260 galaxies. The selection function in the model is as follows: Galaxies had to be within the declination limits of southern HIPASS ($\delta < 2^\circ$) and they had to have a velocity within the HIPASS velocity range of $12\,700 \text{ km s}^{-1}$. The probability of selecting a galaxy that satisfied these conditions was equal to the HIPASS completeness function \citep{Zwaan2004} which depends on the integrated flux and peak flux density.

The galaxies in each of these simulations were binned into bins of equal width in absolute K-band magnitude and their mass spectra stacked. Examples of these stacked spectra can be seen in Fig.~\ref{Plot B1} (left). The widths of these stacked profiles were measured producing $W_{50}^{\rm stack}$ for each bin. In addition to these widths, dispersion-less inclination-corrected individual galaxy profiles were generated and their widths were geometrically averaged to produce $W_{50}^{\rm ref}$ for each bin. Using these widths, we can calculate $\mathcal{F}_\theta$, and thus plot $\mathcal{F}_\theta$ versus $\theta$, the minimum inclination angle of galaxies used in a stack (Fig.~\ref{Plot B5}). From this figure, it can be seen that the sensitivity-limited data sets require a correction factor $\mathcal{F} = 0.935 \pm 0.005$ which agrees with both the analytical galaxies with differential rotation from \S\ref{galaxies with rotation curves}, and the volume-complete sample from \S\ref{Millennium simulated galaxies}. Another important point to note from this figure is, just like Fig.~\ref{Plot B1}, at $\theta = 0$, the scatter between the bins is not systematically linked with magnitude or $s$. Due to $\mathcal{F}$'s similar value to the asymptotic value from equation~(\ref{exp 1}), this equation was used for the correction factor for the TFR.
 
The main difference in the volume-complete data set and this sensitivity-limited data set is the different inclination distributions. The volume-complete sample of galaxies follow a $\sin{i}$ inclination distribution, while by comparison, the HIPASS simulations have an over abundance of face-on galaxies and less edge-on galaxies (shown in Fig.~\ref{Plot B6}). This is due to $S_{\rm peak}$ being a function of inclination. Surprisingly, the HIPASS inclination distribution does not change the value of $\mathcal{F}$ compared to the value calculated when considering data from a $\sin{i}$ distribution.
 
The galaxies in the sensitivity-limited data set were then plotted on a TFR. This was done by first creating observed (inclined and dispersed) spectra for each of the galaxies. The resulting spectra were then stacked within their assigned magnitude bins. Each stack had $W_{50}^{\rm stack}$ measured, and corrected using $\mathcal{F}$ from equation~(\ref{exp 1}). The rotation velocity was calculated from this corrected width ($W_{50}^{\rm stack,c}$) as $W_{50}^{\rm stack,c}/2$. This $V_{\rm rot}$ is the value used -- along with the average magnitude of the galaxies in the stack -- to create a blue data point for Fig.~\ref{Plot B4}.

Whilst creating data points for individual galaxies on the TFR, we use dispersion-less, edge-on rest-frame spectra. We then measure $W_{50}^{\rm ind}$ for each galaxy. The rotation velocity of the galaxy is then calculated as $W_{50}^{\rm ind}/2$ and used, along with the absolute magnitude, to place it on the TFR as a red data point (Fig.~\ref{Plot B4}). Only one realisation of the HIPASS simulation is shown, however, the fits to the data shown are the mean values across all five HIPASS realisations. Similarly, the errors in these parameters, shown as shaded regions, are calculated as the standard error in each parameter across all five realisations. Fits were done using the R package hyper.fit \citep{Robotham2015}. We included the intrinsic scatter as a fitting parameter for the individual galaxies, but not for the stacked data points. The slopes and offsets for the fitting equation

\begin{equation}\label{TFRslope}
	M_{\rm k} = b + a \times \log_{10}{V}
\end{equation}
are summarised in the table of Fig.~\ref{Plot B4}. The difference in the measured slope, $a$, for the stacked and non-stacked data, $\Delta a = 0.05 \pm 0.06$\footnote{Calculated using $\Delta a = a_{1} - a_{2} \pm \sqrt{\sigma_{a_1}^2 + \sigma_{a_2}^2}$ where $\sigma_{a_1}$ is the error in $a_1$ and $\sigma_{a_2}$ is the error in $a_2$.}, is consistent within uncertainties. Similarly, the zero-point offsets, $b$, measured from each method show agreement; $\Delta b = 0.1 \pm 0.1$.

We have established that with a correction value, $\mathcal{F}$, given in equation~(\ref{exp 1}), stacking can reproduce the TFR for a sensitivity-limited sample. We now investigate a sensitivity-limited sample with artificially introduced noise to demonstrate the power of HI stacking when applied to non-detections.

\subsection[]{Gaussian noise}
\label{Gaussian noise}

\begin{figure}
\begin{centering}
\includegraphics[scale=0.45]{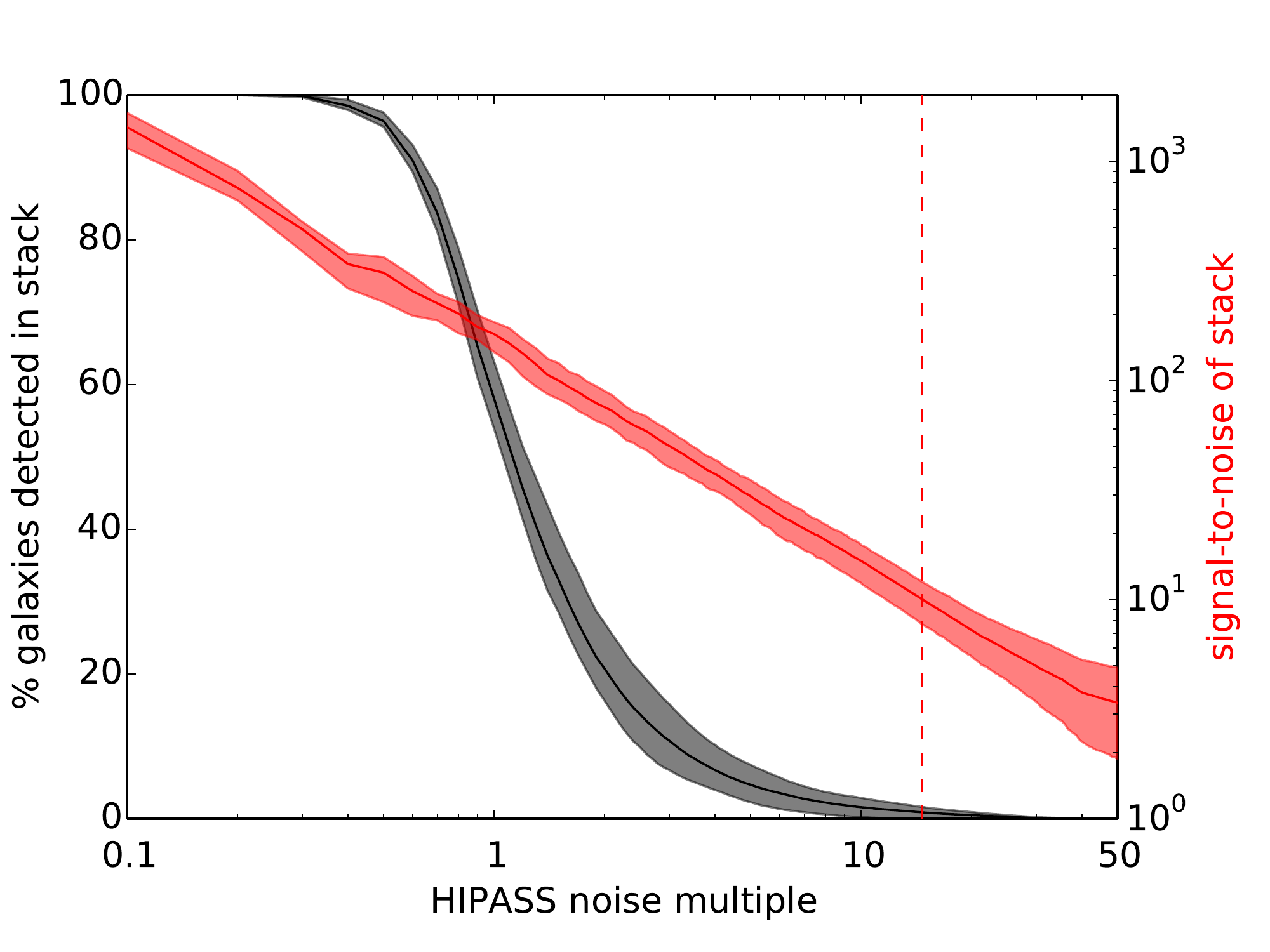}
\caption{Noise is plotted on the x-axis in multiples of the HIPASS noise ($13 \text{mJy beam}^{-1}$) from 0.1$\times$ to 50$\times$ in increments of 0.1$\times$. The black line corresponds to the left axis, showing the mean percentage of galaxies detected in a stack of Milky Way-sized galaxies ($M_k\approx-21$) and one standard deviation from the mean of the five HIPASS volumes shown as a grey shaded region. The red line corresponds to the right axis showing the mean signal-to-noise ratio of stacks containing Milky Way-sized galaxies. One standard deviation from the mean is shown as a red shaded region. The red dotted line intersects at a signal-to-noise of 10 corresponding to the noise used for the Tully-Fisher relation in Fig.~\ref{HIPASS TFR sim with noise} It can also be used to find the percentage of Milky Way-sized galaxies directly detectable in this sample (0.9\%).\label{dual noise plot}}
\par\end{centering}
\end{figure}

\begin{figure}
\begin{centering}
\includegraphics[scale=0.45]{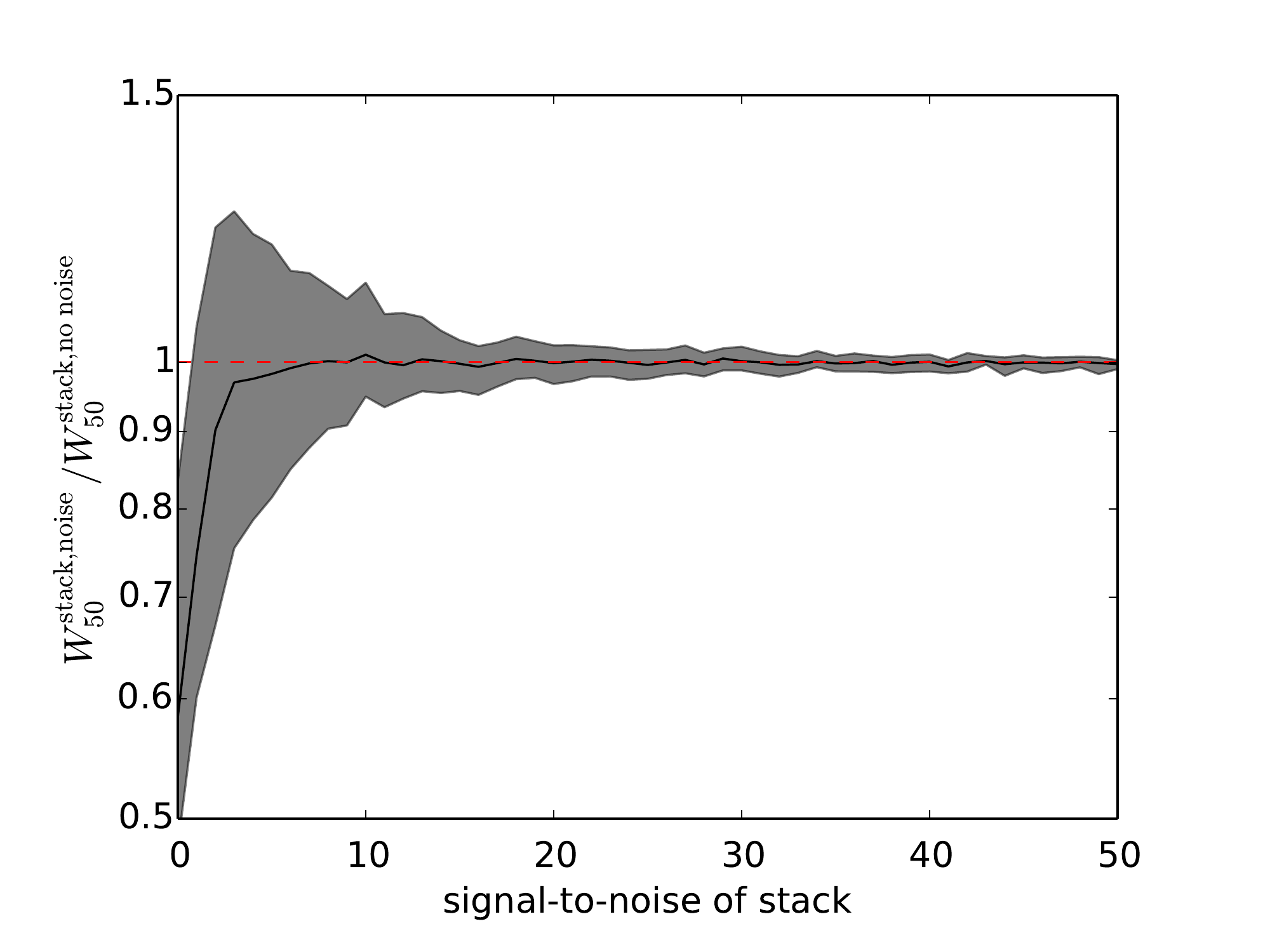}
\caption{The normalised stack width as a function of signal-to-noise in the stack. The grey shaded region shows the standard deviation across all magnitudes and all five simulated HIPASS volumes.\label{noise correction plot}}
\par\end{centering}
\end{figure}

\begin{figure}
\begin{centering}
\includegraphics[scale=0.45]{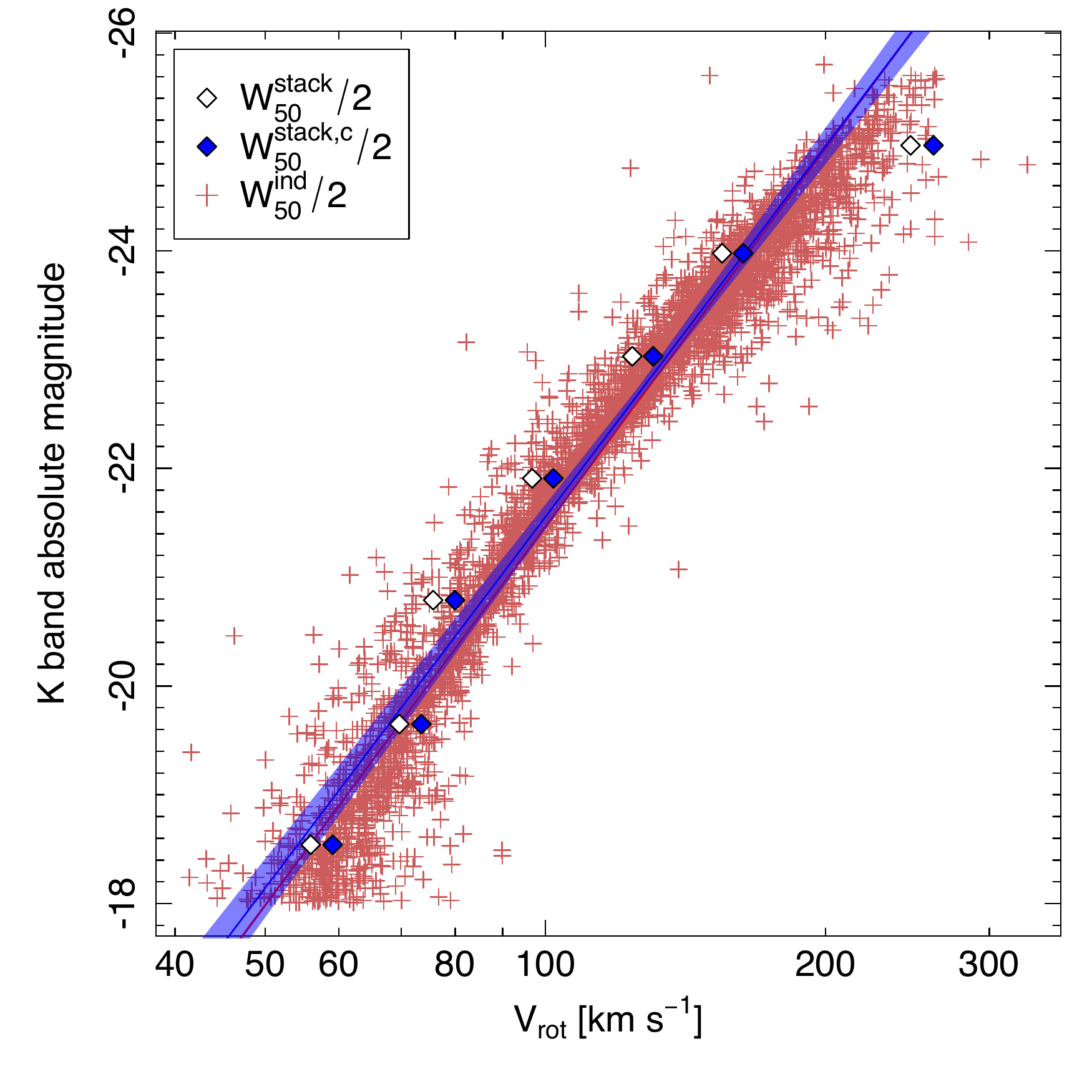}
\caption{The Tully-Fisher relation for a HIPASS simulation using the S$^3$-SAX galaxies with $188.5 \text{ mJy beam}^{-1}$ (14.5$\times$ the HIPASS noise). The red line is the best fit to the red data points (individual galaxy measurements made without noise) across all five HIPASS realisations  The blue line is the mean best fit line for stacks from the five realisations of HIPASS and the blue shaded region represents the error in the slope and offset of this fit from the five realisations.\label{HIPASS TFR sim with noise}}
\par\end{centering}
\end{figure}

Using the same galaxies from section \ref{sensitivity-limited simulated galaxies}, we introduced Gaussian noise into the simulation. We varied this noise to see how well the TFR is recovered from stacks with decreasing signal-to-noise ratios.

To achieve this, we added an equal level of noise to all galaxies in the observers frame. All noise values are presented in multiples of $13 \text{ mJy beam}^{-1}$ to allow for direct comparison with the results from the HIPASS dataset \citep{Meyer2004}. We also removed all resolved galaxies (those with diameters larger than the HIPASS beam). This eliminated extreme outliers, allowing the number of galaxies directly detected prior to stacking to fall to zero over the signal-to-noise range probed. The velocity resolution was dropped to $13 \text{ km s}^{-1}$ to more closely mimic that of HIPASS at $13.2 \text{ km s}^{-1}$ \citep{Meyer2004}.

After each galaxy had the appropriate level of noise added, the noisy {\HI} spectra were stacked in their appropriate magnitude bins. The width of these stacked spectra were measured by fitting the profile described in equations~(\ref{diff rotation stack}) and (\ref{add dispersion}) from section~\ref{galaxies with rotation curves}, leaving the width and height as fitting parameters. $W_{50}^{\rm stack}$ was directly measured from this fitted profile. The signal-to-noise was also measured for each stack. This process was repeated for noise levels from 0.1$\times$ to 50$\times$ the HIPASS noise level in 0.1$\times$ increments.

Fig.~\ref{dual noise plot} shows the percentage of directly detected galaxies as a function of the noise level in the simulation (black). It also shows the signal-to-noise of the stack as a function of the noise in the simulation (red). The data for both the black and red regions were taken from galaxy bins with absolute magnitudes around $-21$. The red dashed line shows the noise level where the mean stack signal-to-noise of the five simulated HIPASS volumes drops to 10, approximately $14.5 \times$ the HIPASS noise.  At this level of noise, despite only $0.9 \pm{+0.9}\%$ of galaxies being detected individually at a $\sim$10$\sigma$ level, we can still reliably recover the width of the stacked spectrum (Fig.~\ref{noise correction plot}) needed for Tully-Fisher purposes (indeed, the mean recovered profile width for the noisy stacks deviates less than 1\% from its noise-free equivalent down to a signal-to-noise ratio of 5, although scatter in the recovered widths does increase substantially below a signal-to-noise of 10).  Fig.~\ref{HIPASS TFR sim with noise} demonstrates this point directly, showing the ability of HI stacking to reliably recover the TFR for the simulated data with $14.5 \times$ the noise of HIPASS.  We note that as such, the results of this simulation indicate that HI stacking could have been used to recover the Tully-Fisher relation from a survey taking just 0.5\% of the observing time used for HIPASS.

The above ability of {\HI} stacking to reliably recover the TFR from noisy {\HI} datasets where the direct detection of sources is difficult also shows the potential of this technique to be applied in other noise-limited regimes, such as the study of the Tully-Fisher relation at higher redshifts or lower {\HI} masses.

We now turn our attention towards the real HIPASS dataset to measure the TFR.

\section[]{HIPASS analysis}
\label{HIPASS analysis}

\begin{figure*}
\begin{centering}
\includegraphics[scale=0.45]{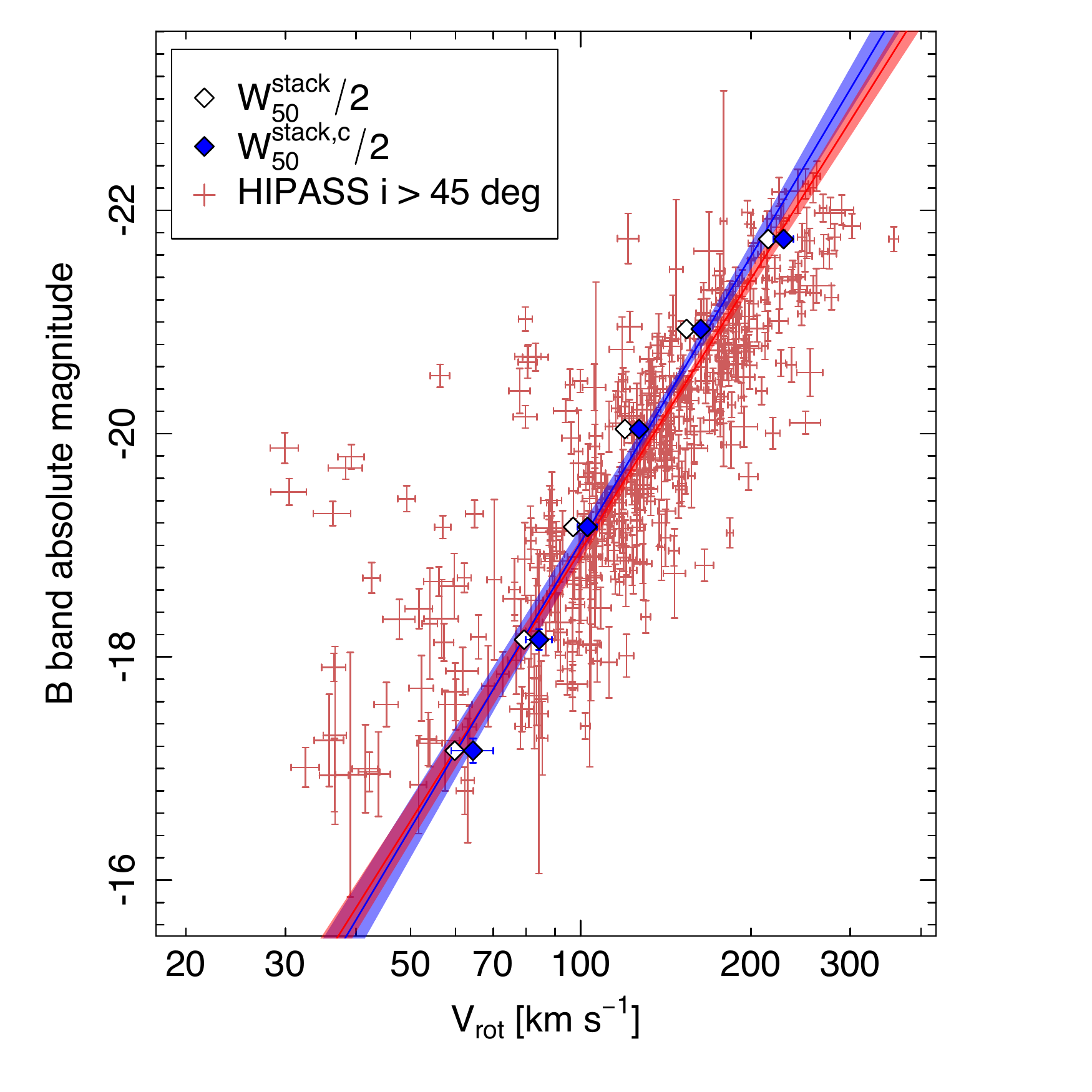}
\includegraphics[scale=0.45]{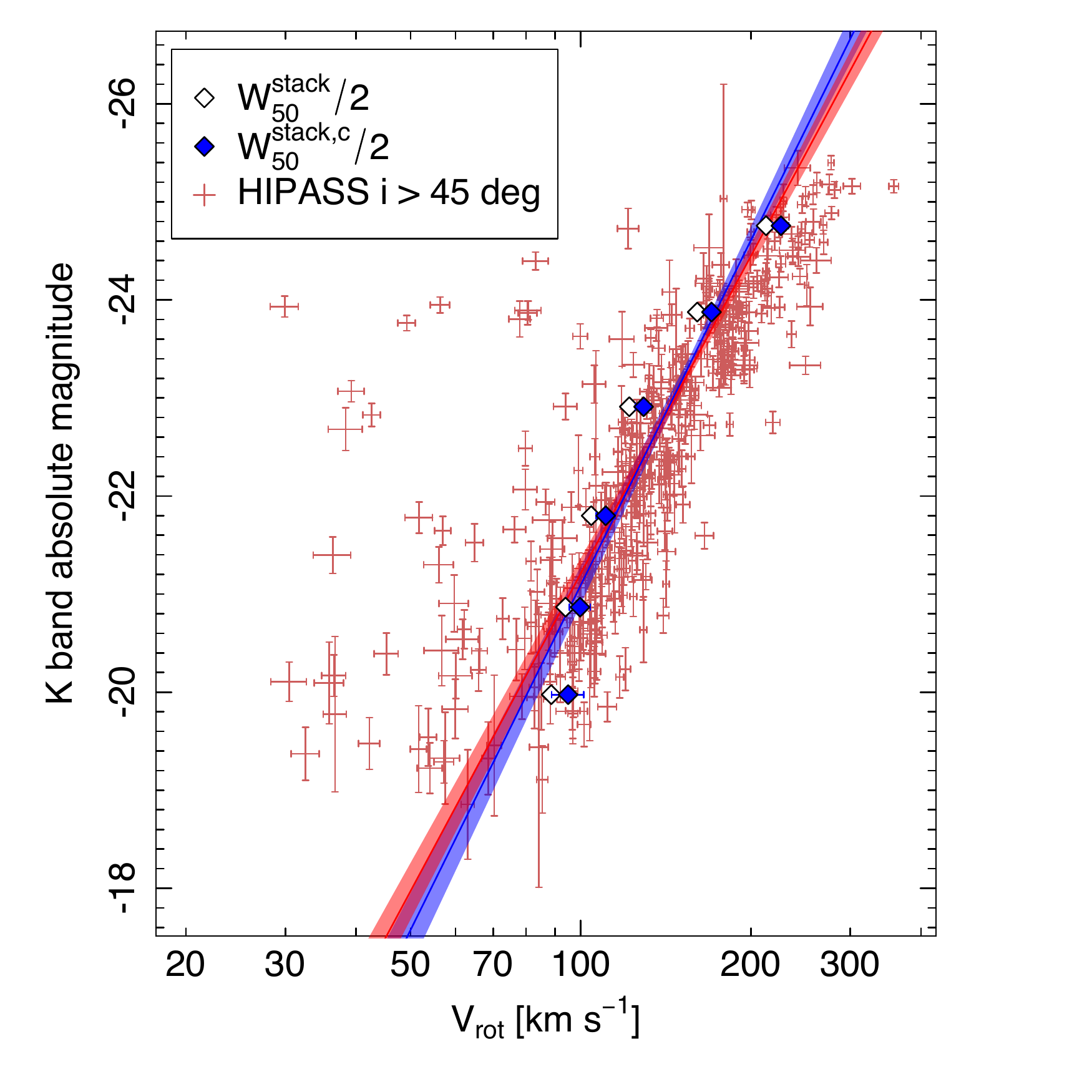}
\begin{tabular}{c c c c c} 
\hline 
Band & Relation & Slope & Offset & \# of galaxies \\ [0.5ex] 
\hline 
B & Stacked & $-8.5 \pm 0.4$ & $-2.0 \pm 0.9$ & $617$ \\
B & Individual & $-8.1 \pm 0.3$ & $-2.8 \pm 0.6$ & $409$ \\
K & Stacked & $-11.7 \pm 0.6$ & $2 \pm 1$ & $591$ \\
K & Individual & $-10.8 \pm 0.4$ & $0.3 \pm 0.9$ & $391$ \\ [1ex]
\hline\hline 
\end{tabular}
\caption{The Tully-Fisher relation in B-band (left) and K-band (right) for both the stacked and individual relations. HIPASS galaxy stacks are plotted as white diamonds on a Tully-Fisher plot. The plot shows average magnitude versus rotation velocity. The corrected stack widths ($W_{50}^{\rm stack,c} = \mathcal{F}^{-1}W_{50}^{\rm stack}$) are also plotted as blue diamonds. Individual galaxies are plotted with errors in red. There is a solid red line fit to the individual galaxies, as well as a blue line for the corrected stack points with the error indicated by the blue and red shaded regions. The slope and offset are of the form in equation~(\ref{TFRslope}).\label{Plot C1}}
\par\end{centering}
\end{figure*}

Finally we test our derived correction factor $\mathcal{F}$ on the HIPASS data set and compare the stacked TFR to the TFR created using the HOPCAT galaxies.

Galaxies were selected from the {\HI} Parkes All-Sky Survey (HIPASS) Catalogue (HICAT) using the same method as \citep{Meyer2008}. HIPASS is a blind {\HI} survey created using the 64 meter radio telescope in Parkes, NSW, Australia. HICAT contains 4315 galaxies from the entire southern sky with declination $\delta < 2$ and velocities in the range v = 300 to 12 700 km s$^{-1}$. For more information about HICAT, see \citet{Meyer2004, Zwaan2004}.

Galaxy positions were found by \citet{Doyle2005} by centring a 15-arcmin SuperCOSMOS image on the HIPASS locations. Overlaid on the image were galaxies found using SE{\sc xtractor} and galaxies in the NASA Extragalactic Database (NED) and 6dFGS. Optical matches were manually chosen from the available galaxies. The resulting galaxy match catalogue is called HOPCAT (the HIPASS Optical Catalogue). Full HOPCAT details can be found in \citet{Doyle2005}. The galaxies in HOPCAT are taken to be the {\it location} of the galaxies, however, for optical magnitudes, the ESO-LV catalogue was used \citep{Lauberts1989}.

Near-infrared galaxy data was gathered from the Two Micron All Sky-Survey (2MASS) Extended Source Catalogue (XSC). This data set covers J (1.11-1.36 $\mu$m), H (1.50-1.80 $\mu$m) and K$_s$ (2.00-2.32 $\mu$m) bands. 2MASS has a 23 arcsec resolution with 1 arcsec pixels. The 1$\sigma$ background noise is 21.4 mag arcsec$^{-2}$ in the J band, 20.6 mag arcsec$^{-2}$ in H and 20.0 mag arcsec$^{-2}$ in $\mathrm{K_s}$. For more information about this data set see \citet{Jarrett2003} and \citet{Cutri2006}. The HOPCAT galaxies were matched to the closest galaxy in the 2MASS catalogue.

Galaxies used for stacking were selected as in \citet{Meyer2008}. Once stacks were created, we cut any stacks with less than 5 galaxies, due to large errors. The galaxies from our comparison sample were also selected as in \citet{Meyer2008}, however a $45^\circ$ inclination cut was introduced, as the error in rotation velocities due to the inclination correction is very large for more face-on galaxies.

All galaxies with ESO-LV optical magnitudes were separated into equally spaced magnitude bins and stacked. Each bin was chosen to be roughly 1 magnitude in width to match all the previous work. We measured $W_{50}^{\rm stack}$ for each of the stacked spectra and corrected them by the correction factor ($\mathcal{F}$) calculated from equation~(\ref{exp 1}). In the table accompanying Fig.~\ref{Plot C1} we show the slope and offset calculated from our two data sets using the hyper.fit package in R \citep{Robotham2015}.

The B-band stacked relation agrees well with the individual galaxy relation with a slope difference of $\Delta a = 0.5 \pm 0.5$ and a zero-point offset difference of $\Delta b = 0.8 \pm 1.1$ when comparing the two techniques.

The K-band stacked relation is in poorer agreement with the fit to the corresponding K-band comparison sample with $\Delta a = 0.9 \pm 0.8$ and $\Delta b = 2 \pm 2$, but they still agree with one another.

The stacked relations reproduce the TFR parameters well, although with up to 50\% larger errors for this data set.

The HIPASS galaxies included in our stacks have a different inclination distribution to the assumed $\sin{i}$ function. Although the stack widths appear to be robust to inclination distribution, as shown in \S\ref{sensitivity-limited simulated galaxies}, it may have had a more significant effect on this smaller data set.

\section[]{Conclusion}
\label{conclusion}

The goal of this paper is to see if the same TFR is recovered using an {\HI} stacking method than when using each galaxy individually. To that end, we stack progressively  more realistic galaxies, the results of which are summarised below.

\vspace{-3mm}
\subsection*{Identical mock galaxies}

We create mock galaxies with constant and identical circular velocity ({\it not} solid rotators) and differential circular velocity (equations~\ref{const rot} \& \ref{diff rotation}) seen under random inclinations and smoothed by a dispersive component. Stacking the {\HI} emission lines of these simple galaxies (equations~\ref{stackeqn} \& \ref{diff rotation stack}) allows us insight into the measurement of rotation velocities from stacked line profiles.

\vspace{-3mm}
\subsubsection*{Constant circular rotation}

\begin{myitemize}
\item We find that the width of a stacked {\HI} line profile $W_{50}^{\rm stack}$ is exactly identical to the width of the non-dispersed, edge-on profiles $W_{50}^{\rm ref}$ of the individual galaxies; hence $\mathcal{F} = W_{50}^{\rm stack}/W_{50}^{\rm ref} = 1$ in this simplistic model.
\item We find that $\mathcal{F}$ is robust to the magnitude of gas dispersion included in our simulated galaxy.
\item We show that average rotation velocity can be recovered from a sample of galaxies without ever needing to measure the inclination angle of the galaxies.
\end{myitemize}

\vspace{-8mm}
\subsubsection*{Differential circular rotation}

\begin{myitemize}
\item Upon adopting a more accurate model for the {\HI} disk, with differential (linear) rotation, the width of a stacked {\HI} line, again composed of galaxies with identical edge-on {\HI} profiles, becomes slightly smaller, such that $\mathcal{F} = 0.95 \pm 0.01$.
\end{myitemize}

\vspace{-8mm}
\subsection*{Simulated galaxies}

Using a volume-complete sample of simulated galaxies from the S$^3$-SAX simulation gives us a more realistic set of galaxies, and the sensitivity-limited subsample mimics the HIPASS selection function to give us insight into how our method behaves with even more realistic data sets. The volume-complete and sensitivity-limited simulations differ from the above identical mock galaxies as all the galaxies used now have different edge-on profiles.

\vspace{-3mm}
\subsubsection*{Volume-complete}

\begin{myitemize}
\item We create stacks of all galaxies in K-band bins that are 1 magnitude wide and measure a value of $\mathcal{F} = 0.93 \pm 0.01$ for the volume-complete simulation. This is very close to the value of $\mathcal{F} = 0.95 \pm 0.01$ we measured from the galaxies with differential rotation velocity profiles.
\item The effect of including or excluding non-spiral galaxies into our samples makes little ($<1\%$) difference to the measured $W_{50}^{\rm stack}$.
\end{myitemize}

\vspace{-7mm}
\subsubsection*{Sensitivity-limited}

\begin{myitemize}
\item We found from the sensitivity-limited subsample of the S$^3$-SAX simulation (mimicking the HIPASS selection) that the correction factor given in equation~(\ref{exp 1}) works well for stacks with a few hundred galaxies.
\item We also showed that the standard method of deriving the TFR from individual galaxies and our stacking method agree with each other for five independent data sets.
\end{myitemize}

\subsubsection*{Gaussian noise}
\begin{myitemize}
\item We have shown that by increasing the noise in the sensitivity-limited subsamples of the S$^3$-SAX simulation, we can reliably recover the TFR using stacking at a noise level where less than 1\% of the galaxies are detected individually.
\item Of particular note; this noise level is equivalent to a reduction in observation time by up to 99.5\%.
\end{myitemize}

\vspace{-7mm}
\subsection*{HIPASS}

The stacking method used to derive the TFR was compared to the standard method used in \citet{Meyer2008} using individual galaxies.

\begin{myitemize}
\vspace{-3mm}
\item In the B-band, the stacked relation matches quite well with the relation derived from a galaxy-by-galaxy analysis.
\item The stacked TFR is in poorer agreement in the K-band, but still within errors of the galaxy-by-galaxy analysis.
\end{myitemize}

\vspace{-3mm}
We show that with an {\HI} selected data set and no knowledge of the individual galaxy inclinations, or even the inclination distribution of the data, the TFR can be recovered via the spectral stacking technique we investigated.

In our next publication we will extend this further by extracting radio data centred on the 6dFGS galaxies and testing the stacking technique on a sample of non-detections for potential application of this technique at higher redshifts or lower masses.

\section*{Acknowledgments}

We are grateful to the S$^3$-SAX and HIPASS collaborations. SM wishes to thank Stefan Westerlund, Laura Hoppmann, Jacinta Delhaize, Matthew Pearce, Rebecca Lange and Angus Wright for coding assistance and Taylem Frost for editorial work. We acknowledge use of the \textsc{topcat} software package (http://www.star.bristol.ac.uk/{\textasciitilde}mbt/topcat/) and \textsc{miriad} \citep{Sault1995} for data reduction. This research was conducted by the Australian Research Council Centre of Excellence for All-sky Astrophysics (CAASTRO), through project number CE110001020.

\bibliographystyle{mn2e}
\bibliography{scottsbibliography}

\appendix
\section{Solution for isotropic stack of the basic model}

We are interested in an analytical closed-form solution of equation~(\ref{stackeqn}) describing the case of an isotropic stack of idealised emission lines from flat, axially symmetric, transparent disk galaxies with circular orbits at a constant velocity $\vmax$ and no dispersion. To solve this equation, let us first remember that such a stacked emission line comes about when observing an equal amount of material flying in every direction at a fixed velocity $\vmax$. Therefore the situation is equivalent to observing a single spherical shell of uniform surface density expanding at $\vmax$. Because the observer only sees the velocity component along the line-of-sight, the profile $\rho_{\rm const}^{\rm stack}(v)$ is then equal to the 1D-density profile resulting from projecting the surface of a unit-sphere onto a straight line. Upon parametrising this surface in spherical coordinates with longitude $\phi\in[0,2\pi]$ and latitude $\theta\in[-\pi/2,\pi/2]$ such that $v=\sin\theta$, we obtain
\begin{equation}
\begin{split}
	\rho_{\rm const}^{\rm stack}(v) &= \frac{1}{4\pi}\int_0^{2\pi}\!\!{\rm d}\phi\cos\theta\,\frac{{\rm d}\theta}{{\rm d}v} \\
						        &= \frac{1}{2}\cos(\arcsin v)\,\frac{{\rm d}\theta}{{\rm d}v} \\
						        &= \frac{1}{2}\frac{\sqrt{1-v^2}}{\sqrt{1-v^2}}=\frac{1}{2}.
\end{split}
\end{equation}
This equation is valid on the interval $v\in[-1,1]$ covered by the projection of the unit sphere. Outside this interval the projection vanishes, hence $\rho_{\rm const}^{\rm stack}(v)=0$. In conclusion, the profile of an isotropic stack of galaxies rotating at a constant velocity $\vmax$ is a top-hat bounded between $V\in[-\vmax,\vmax]$.

\section{Derivation of edge-on emission line profiles for differential circular velocity}

To derive the emission line profile of an edge-on galaxy with differential circular velocity (equation~\ref{diff rotation}), we start with the equation describing a line profile for an edge-on disk with constant circular velocity (equation~\ref{const rot}). We then change the constant velocity profile to
\begin{equation}\label{Vreq}
	V(r) = V_{\rm max} \left(1 - e^{-r / r_{\rm flat}}\right).
\end{equation}
Since this velocity is now a function of $r$, the distribution of mass within the disk now affects the shape of the emission line. We use a simple exponential surface density given by
\begin{equation}
	\Sigma_{\rm HI}(r) = \frac{1}{2\pi r_{\rm HI}^2} e^{-r/r_{\rm HI}},
\end{equation}
when normalised to the {\HI} mass. Replacing $V_{\rm max}$ in equation~(\ref{const rot}) with equation~(\ref{Vreq}) and integrating over mass, we get
\begin{equation}\label{B3}
	\rho_{\rm diff}^{\rm edge}(v) = \int_0^\infty {\rm d}M(r) \frac{1}{\pi \sqrt{\left[\frac{V(r)}{V_{\rm max}}\right]^2 - v^2}},
\end{equation}
where ${\rm d}M(r) = {\rm d}A ~ \Sigma_{\rm HI}(r) = r {\rm d}r \frac{e^{-r/r_{\rm HI}}}{r_{\rm HI}^2}$. Thus,
\begin{equation}\label{B4}
	\rho_{\rm diff}^{\rm edge}(v) = \int_0^\infty \frac{r {\rm d}r}{r_{\rm HI}^2} \frac{e^{-r/r_{\rm HI}}}{\pi \sqrt{\left[1 - e^{-r/r_{\rm flat}}\right]^2 - v^2}}.
\end{equation}
Using the substitution $r' = r/r_{\rm HI}$ and $r_{\rm HI}/r_{\rm flat} = 3$, consistent with the regular disks in the THINGS catalogue \citep{Leroy2008}, into equation~(\ref{B4}), we finally end up with
\begin{equation}
	\rho_{\rm diff}^{\rm edge}(v) = \int_0^\infty \!\!{\rm d}r' \frac{r' e^{-r'}}{\pi \sqrt{\left[1 - e^{-3r'}\right]^2 - v^2}}.
\end{equation}
%

\bsp

\label{lastpage}

\end{document}